\documentclass[12pt,a4paper]{article}

\usepackage[left=1.2in, right=1.2in]{geometry}

\usepackage{amssymb}
\usepackage{amsfonts}
\usepackage{amsmath}
\usepackage{graphicx}
\usepackage[font=small]{caption}
\usepackage{subcaption}
\usepackage{bm}
\usepackage{algorithm}% http://ctan.org/pkg/algorithms
\usepackage{algpseudocode}% http://ctan.org/pkg/algorithmicx
\usepackage{varwidth}
\usepackage{placeins}

\newcommand{\comment}[1]{}  %comment not showed
\newcommand{\qed}{\rule{1.5mm}{2mm}\vspace{0.1in}}

\newcommand{\ignore}[1]{}

\algtext*{EndFor}% Remove "end for" text
\algtext*{EndIf}% Remove "end if" text

\title{A ``Quantal Regret'' Method for Structural Econometrics in Repeated Games}

\author{
Noam Nisan\thanks{Rachel \& Selim Benin School of Computer Science \& Engineering and Federmann Center for the Study of Rationality, The Hebrew University of Jerusalem, Israel; and Microsoft Research.  Supported by ISF grant 1435/14 administered by the Israeli Academy of Sciences and by
Israel-USA Bi-national Science Foundation (BSF) grant number 2014389.}
\and
Gali Noti\thanks{Rachel \& Selim Benin School of Computer Science \& Engineering and Federmann Center for the Study of Rationality, The Hebrew University of Jerusalem, Israel.  Supported by the Adams Fellowship Program of the Israel Academy of Sciences and Humanities. }
}

\begin{document}
\maketitle

\begin{abstract}
We suggest a general method for inferring players' values from their actions in repeated games.  
%The method improves upon
%the recent suggestion of \cite{Eva} that is based on 
%the assumption that players minimize their regret, 
%by taking instead 
%a ``quantal'' version 
The method extends and improves upon the recent
suggestion of (Nekipelov et al., EC 2015)  
and is based on the assumption that players are 
{\em more likely} to exhibit sequences of actions that
have {\em lower regret}.

We evaluate this ``quantal regret'' method 
on two different datasets from experiments of 
repeated games with controlled player values: 
those of (Selten and Chmura, AER 2008) on a variety 
of two-player 2x2 games and our own experiment on ad-auctions (Noti et al., WWW 2014).  
We find that the quantal-regret method is consistently and significantly 
more precise than either ``classic'' econometric methods 
that are based on Nash equilibria, 
or the ``min-regret'' method of 
(Nekipelov et al., EC 2015).
\end{abstract}

\sloppy

\section{Introduction} \label{sec:intro}

%intro.tex

\FloatBarrier

\subsection*{Motivation}

Consider some on-line computational platform where users interact repeatedly in some strategic context.  Natural
examples abound including ad-auctions, dating sites, or dynamic allocation of cloud resources.  In such settings
human players are interacting in some repeated strategic game that is defined by the platform.  The on-line platform 
observes the players'
actions in this game (e.g., a sequence of bids in an auction), but does not have access to their private information
(e.g., players' values in the auction).  This latter information is what is really important to the designer and 
to the users
of the platform, as the goals and the level of success of the platform are really defined
in terms of this ``real world'' underlying information.  
Thus, reliable estimates of this private information are essential for the platform designer to evaluate how well 
does his platform do, or how may it further be improved.
For example, the identification of cases where 
the revenue (or welfare) of an on-line auction platform may be improved 
requires reliable estimates of the players' true values in these cases.
Can we reliably estimate the unknown private information of participants 
from their observed actions?

This challenge has received specific attention in the case of ad-auctions: \cite{Varian2007,Athey2010} invoke ``classic''
econometric methods assuming that players are close to an {\em equilibrium}, and as the
equilibrium is a function of the private values of the players, they compute in the inverse direction and 
deduce the private values from the observed equilibrium.  This econometric assumption that players reach equilibrium
is rather problematic: first, it is often quite unclear {\em which} equilibrium to assume: Pure Nash? 
mixed-Nash? Which one of the multiple ones?  Correlated?  Bayes-Nash? Which prior?  Even more
importantly, do we really expect humans to reach a mathematically-defined equilibrium?  A much
more robust approach was suggested in \cite{Nekipelov2015}: they instead assume that players are able to (almost)
{\em minimize their regret}.  Formally, this is a weaker assumption since whenever players reach an equilibrium 
(any Nash, correlated, or even coarse-correlated one)
they also minimize their regret.  Furthermore, this assumption seems more plausible from a human
behavior perspective.  In a previous paper \cite{NN2017}, we evaluated this ``min-regret'' econometric method
on data from an ad-auction experiment that we have previously performed \cite{NNY2014}, and found that it was
indeed at least as good as the previous ``equilibrium-based'' methods.  
Unfortunately there was only a minor --
usually not statistically significant -- improvement in the rather low estimation precision. 

In this paper we propose a new method for this estimation, one that does not assume that players 
exactly minimize their regret but rather only that they are {\em more likely} to behave in less regretful ways.  This
is similar in spirit to ``quantal response'' modeling of human behavior in games \cite{QRE1995}, but is applied to the 
regret in the repeated game rather
than to utilities in a single game.  The method is %quite 
general and easy to implement, and should 
apply to most computerized repeated game settings. We evaluate our method -- and compare it to previous methods -- not only
on data from our ad-auction experiment, but also on data from an experiment of
\cite{Selten2008} on repeated two-player 2x2 games (a context in which we are not aware of any previous 
work on this estimation task.)  

Let us start by demonstrating our method (and previous ones) on a simple example.

\subsection*{An Example}
\begin{figure}[t] % slide 12
\begin{subfigure}{.32\textwidth} 
  \includegraphics[scale=0.61]{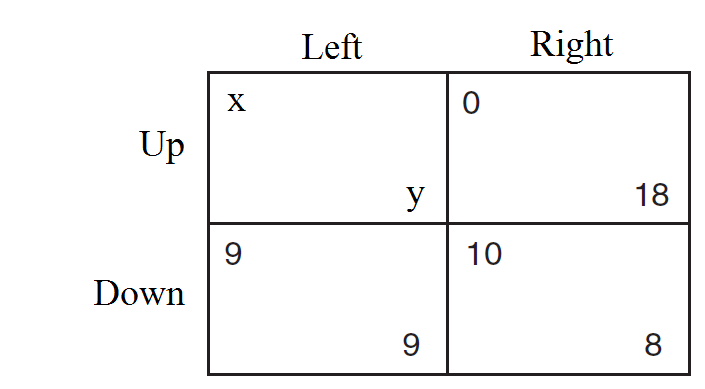}
  \caption{\label{fig:example-mtx-x-y}}
\end{subfigure}
\begin{subfigure}{.32\textwidth}
  \includegraphics[scale=0.61]{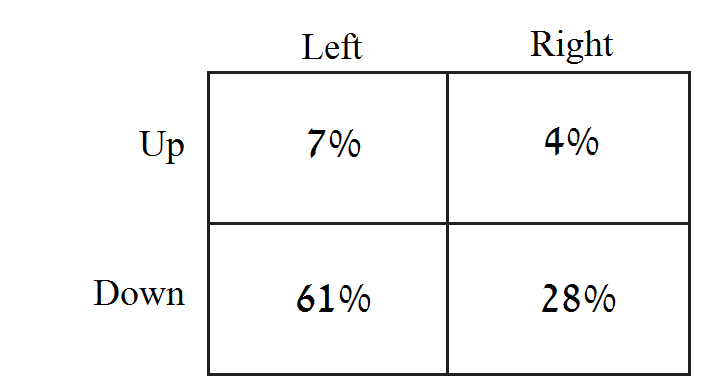}
  \caption{\label{fig:example-freq}}
\end{subfigure}
\begin{subfigure}{.32\textwidth}
  \includegraphics[scale=0.1]{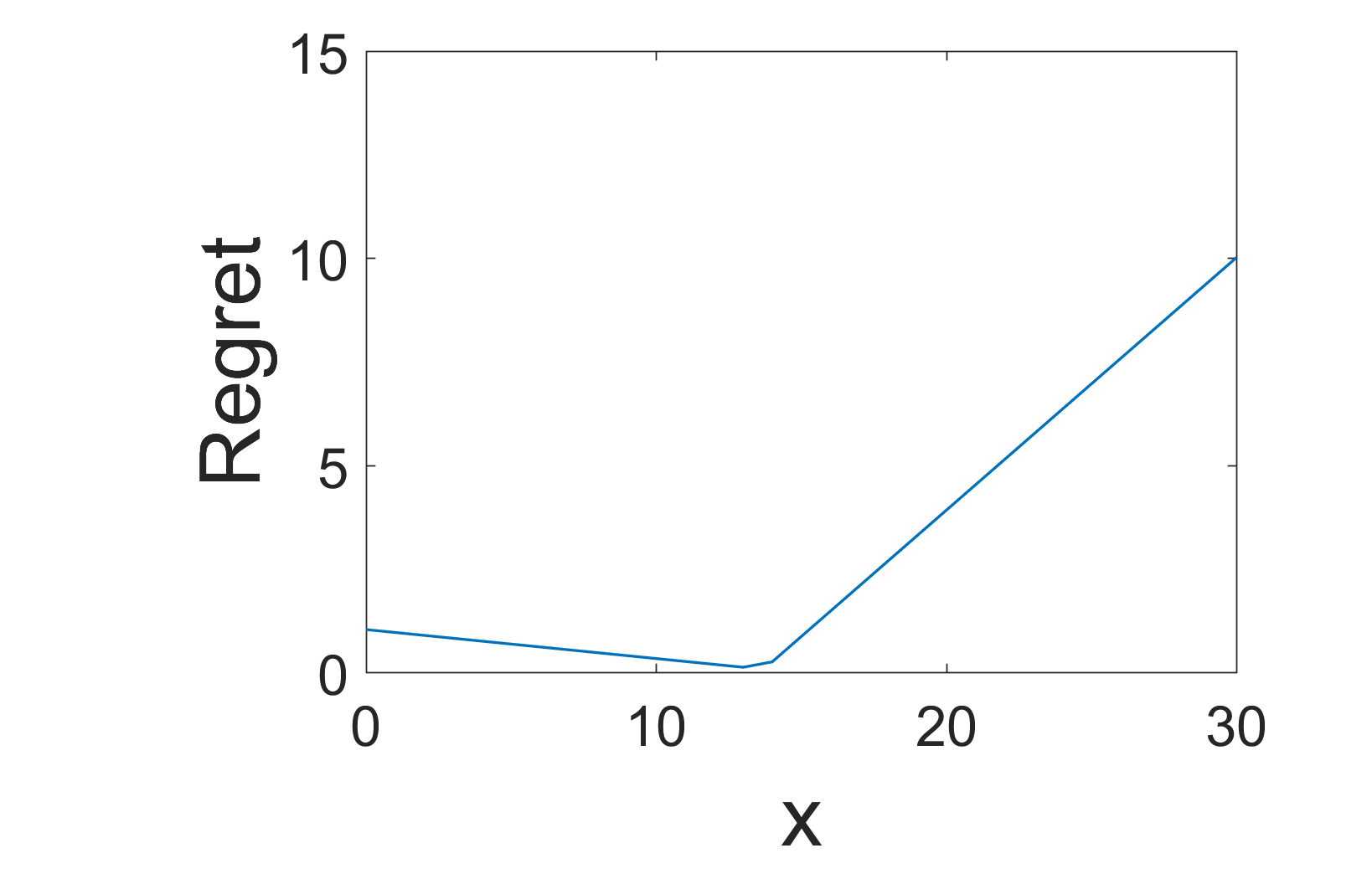}
  \caption{\label{fig:example-regret-curve}}
\end{subfigure}
\caption{An example of the estimation task in a 2x2 game. (\ref{fig:example-mtx-x-y}) The payoff matrix of Game 1 in \cite{Selten2008}, with the parameters that we wish to estimate hidden by $x$ and $y$. 
The upper-left and the lower-right corners in each cell are the payoffs of the row and the column players, respectively. 
(\ref{fig:example-freq}) The average empirical frequency that was obtained in one of the sessions by human players playing Game 1 repeatedly for 200 periods. 
(\ref{fig:example-regret-curve}) The regret curve for the row player in the example, as a function of the parameter $x$. 
\label{fig:example}}
\end{figure}

Consider the following 2x2 two-player game in figure \ref{fig:example-mtx-x-y}, 
where we are only given six of the eight parameters defining the game, and
need to infer the unknown ones, $x$ and $y$, from empirical data on players' actions
when repeatedly playing the game.  This game (with some specific values of $x$ and $y$) is one of those that \cite{Selten2008} have run in experiments with human subjects, and the empirical frequency of play
in one of their sessions is given in figure \ref{fig:example-freq}. 
What would your prediction of the missing parameter $x$ be?  

A {\em classic equilibrium-based approach} would be to assume that the players reach 
(approximately) a mixed Nash equilibrium of the game. In the single mixed equilibrium of
the game, the
column player must be playing $Left$ with a probability $p=10/(x+1)$, so we can estimate
$\hat{x} = 10/p-1 = 10/(0.07+0.61)-1 \approx 13.7$.\footnote{Note that since $p \le 1$ we must have 
$x \ge 9$, and indeed otherwise $Up$ is dominated by $Down$.}

An adaptation of the {\em regret based approach} suggested by \cite{Nekipelov2015} %in the context of the GSP auction, 
would assume that players (nearly) minimize their regret.  Under this assumption 
it would calculate for every possible value of $x$, 
what {\em would have been the regret} of the row player, {\em had the missing parameter been} $x$.  
The regret 
here is defined as the difference between 
the utility that could have been obtained using the best fixed strategy in hindsight (and
the other players still behaving as they did empirically) and 
the utility that was obtained under the empirical play. 
For example, for $x=13$ the row player's empirical utility is 
$0.07 \cdot 13 + 0.04 \cdot 0 + 0.61 \cdot 9 + 0.28 \cdot 10 = 9.20$, while had he always played $Down$ his utility 
would have been $(0.07+0.61) \cdot 9 + (0.04 + 0.28) \cdot 10 = 9.32$ and had he always played $Up$  
his utility would have
been $(0.07+0.61) \cdot 13 +  (0.04 + 0.28) \cdot 0 = 8.84$. Thus the row player's regret is 
$max(9.32, 8.84)-9.20=0.12$.  
%Similarly, 
More generally, one may calculate the regret of the row player for every possible value of $x$ (in some grid in some valuation range), by: 
%The regret of the row player 
$regret(x)=max(util_{Up}(x),util_{Down}(x))-util_{Emp}(x)$. 
The regret of the row player as a function of the ``hidden value'' $x$
is graphed in Figure \ref{fig:example-regret-curve}.  
The basic ``min-regret'' method would take as its estimate $\hat{x}$ the 
value 
with the lowest regret,\footnote{The specific implementation in \cite{Nekipelov2015} actually looked for the lowest 
{\em relative} regret, but as we demonstrate in Appendix \ref{app:mr-rel-regret}, using absolute regret
gives better estimates, so in the rest of the paper we use the simpler and more precise absolute regret method as a tougher benchmark for comparison.}
which happens to be $13$.\footnote{  
Here we are assuming that possible values are only integers; it is not difficult to see that the lowest regret among all possible real-valued $x$ is obtained at the same point of the equilibrium estimate above, 13.7, as is the case in all 2x2 games (but not generally).  As %implied by 
can be seen in 
Table \ref{fig:bottom-games}, the restriction to integer values gives more precise 
estimates for 2x2 games, so, again, we use the more precise method as a tougher benchmark for comparison.}

Our suggested {\em quantal regret method} 
would first take into account that humans are never exact optimizers. 
It would then 
%would 
look at the regret graph (\ref{fig:example-regret-curve}) and notice that it is much steeper to the right of the min-regret point than it is to its left. 
%It may thus 
Thus, it may 
seem more likely that the real value of $x$ is lower than 13 (where the player only loses a bit from acting the way he did) than that it is higher than 13 (where the
player loses a lot).  We may utilize this observation by
taking a {\em weighted average of the possible values of $x$, where the weights are decreasing with the regret}. Specifically, we take exponentially decreasing weights, as follows: 
$\hat{x} = (\sum_x e^{-\lambda\cdot regret(x)} \cdot x) / (\sum_x e^{-\lambda\cdot regret(x)})$.  For a value of (say) $\lambda = 3$ %5$ 
this would evaluate\footnote{ 
The summation in this example is over integers in the range $0..100$, although due to the exponential decay, values over 20 or so have essentially no effect on the result.}  
to $\hat{x} \approx 10.2$. 
As the value of the ``regret aversion'' constant $\lambda$ grows to infinity, the quantal regret estimate approaches the min-regret estimate.  

For the curious reader let us reveal that
in the actual experiment of \cite{Selten2008} the value taken was $x=10$, and indeed in this specific instance the estimate of the quantal regret method was significantly closer to reality than
that of the min-regret method of \cite{Nekipelov2015} which in turn was slightly better than the ``classic'' method.  As we will show below, this is the usual state of affairs in our data: the quantal regret method
consistently and significantly outperforms the min-regret method which in turn somewhat outperforms classic methods in the two scenarios of repeated games that we have looked at.  

\subsection*{Experimental Results}

%\noindent 
%\paragraph{{\bf Experimental Results:}}
We have evaluated the qunatal-regret method by computing its error on data obtained 
in two different experiments of repeated games. In both cases, a game was
designed by an experimenter who controlled all of its parameters, and observed humans
playing the game repeatedly.  We ``hid from ourselves'' a basic parameter of the game $x$,
and tried to infer this hidden parameter using the observed data.  %(This was done for each parameter of the game, one at a time.)
We then ``un-hid'' the actual value of this estimated parameter
and compared this real value with our estimate. %(in each of the estimation methods that we used). 

We compared three basic econometric methods:
a classical one that assumes that players are in an equilibrium (EQ), the min-regret (MR) 
method suggested in \cite{Nekipelov2015} and our proposed quantal regret (QR) method.  
Our basic comparison
metric is the root mean square error (RMSE) over the estimation errors in a specific setting.
Specifically, the estimation error of an estimate $\hat{x}$ for a parameter whose true value is $x$ is the estimate's distance from this true value, i.e.,  $error(\hat{x})=|\hat{x} - x|$, 
and the RMSE of a set of estimates $S$ is $RMSE(S) = \sqrt{\frac{1}{|S|} \sum_{\hat{x}\in S} error^2(\hat{x})}$. 
We also look at the average of all estimation errors, as well as at the ``$\pm K$ hit-rate'', which is the fraction of estimates that are within an interval of $K$ from the true value. 
%Our main measure for the quality of the estimation methods is the RMSE. 

\begin{table}
\centering
\begin{subtable}{.5\textwidth}
\centering

%table material: games
\resizebox{\textwidth}{!}{ %scale down table to the textwidth
\begin{tabular}{|l|l|l|l|}
\hline
\multicolumn{4}{|c|}{2x2 Games -- Over All Sessions} \\ \hline
	& EQ & MR & QR \\ \hline
RMSE  & 3.41 & 3.25 & 2.29 \\
Average Error  & 2.99 & 2.84 & 2.04 \\
$\pm 3$ Hit Rate  & 68.87\% & 75.00\% & 81.60\% \\
\hline
\end{tabular}
}
\caption{\label{fig:bottom-games}}
\end{subtable} 

\begin{subtable}{1.0\textwidth}
\centering

%table material: auctions
\resizebox{\textwidth}{!}{ %scale down table to the textwidth
\begin{tabular}{|l|l|l|l|l|l|l|l|}
\hline
& \multicolumn{3}{|c|}{Ad auctions -- VCG Sessions}  & \multicolumn{4 }{|c|}{Ad auctions -- GSP Sessions} \\ \hline
	& EQ & MR & QR & EQ1 & EQ2 & MR & QR \\ \hline
RMSE  & 6.46 & 6.26 & 4.22 & 9.73 & 9.87 & 8.02 & 5.09 \\
Average Error  & 5.38 & 5.13 & 3.42 & 7.74 & 8.02 & 6.32 & 3.85 \\
$\pm 6$ Hit Rate  & 61.67\% & 63.33\% & 81.67\% & 48.33\% & 41.67\% & 56.67\% & 81.67\% \\ 
\hline
\end{tabular}
}

\caption{\label{fig:bottom-auctions}}
\end{subtable}

\caption{Estimation results of our proposed quantal regret (QR) method, 
the basic min-regret (MR) method suggested by \cite{Nekipelov2015}, 
and the classic equilibrium-based (EQ) method,  
for two different datasets: 
(\ref{fig:bottom-games}) over all 108 sessions of the 2x2 game dataset; and (\ref{fig:bottom-auctions}) either for VCG or for GSP sessions of the ad auction dataset. In the GSP setting there are two different equilibrium-based methods (EQ1 of \cite{Varian2007} and EQ2 of \cite{Athey2010}).}
\end{table}

Our first dataset is from \cite{Selten2008}, where human subjects were given a two-player 2x2 game and played it for 200 turns.\footnote{The exact
setup is a bit more complicated as the players were split into ``groups'' with 4 row players and 4 column players who were re-matched at random
with each other for each of the 200 periods.  See the details in Section \ref{sec:games}.}  
The data that we have is the empirical frequency of play of each of the four 
possible strategy profiles, $(Up, Left)$, $(Up,Right)$, $(Down,Left)$, $(Down,Right)$,  
during these 200 plays of the game -- exactly like in the example
above.  %We have data for 12 different games, 
12 different games were investigated, 
half of which are constant-sum, 
where for each game we have data from multiple independent sessions: 12 sessions for each of the constant-sum games and 6 sessions for each of the non-constant sum games; for a total of 108 sessions.

For each of these 108 sessions we estimated (separately) each of the 8 parameters (payoffs) defining the game, using each of the three methods.  
%The setup as well as our analysis are given in section \ref{sec:games}.  
The bottom line, as shown in Table \ref{fig:bottom-games},
is that the quantal regret method has a large and statistically significant advantage over the other two methods, 
while the min-regret method is just slightly better than the Nash equilibrium method.\footnote{As mentioned previously, 
for this case of 2x2 games, the variant of the min-regret estimator that allows real values
would be identical to the Nash equilibrium method.  The min-regret method reported here allows only integer values
and gives slightly better results since it usually manages to round in the ``correct direction'' where the regret rises less steeply.} 
%The quality gap between the results is statistically significant (using test XXX, with n=108 independent sessions, looking at the average-square errors of the 8 parameters in each session.)  

\begin{figure}[t] % slide 12
\begin{subfigure}{.32\textwidth} 
  \includegraphics[scale=0.13]{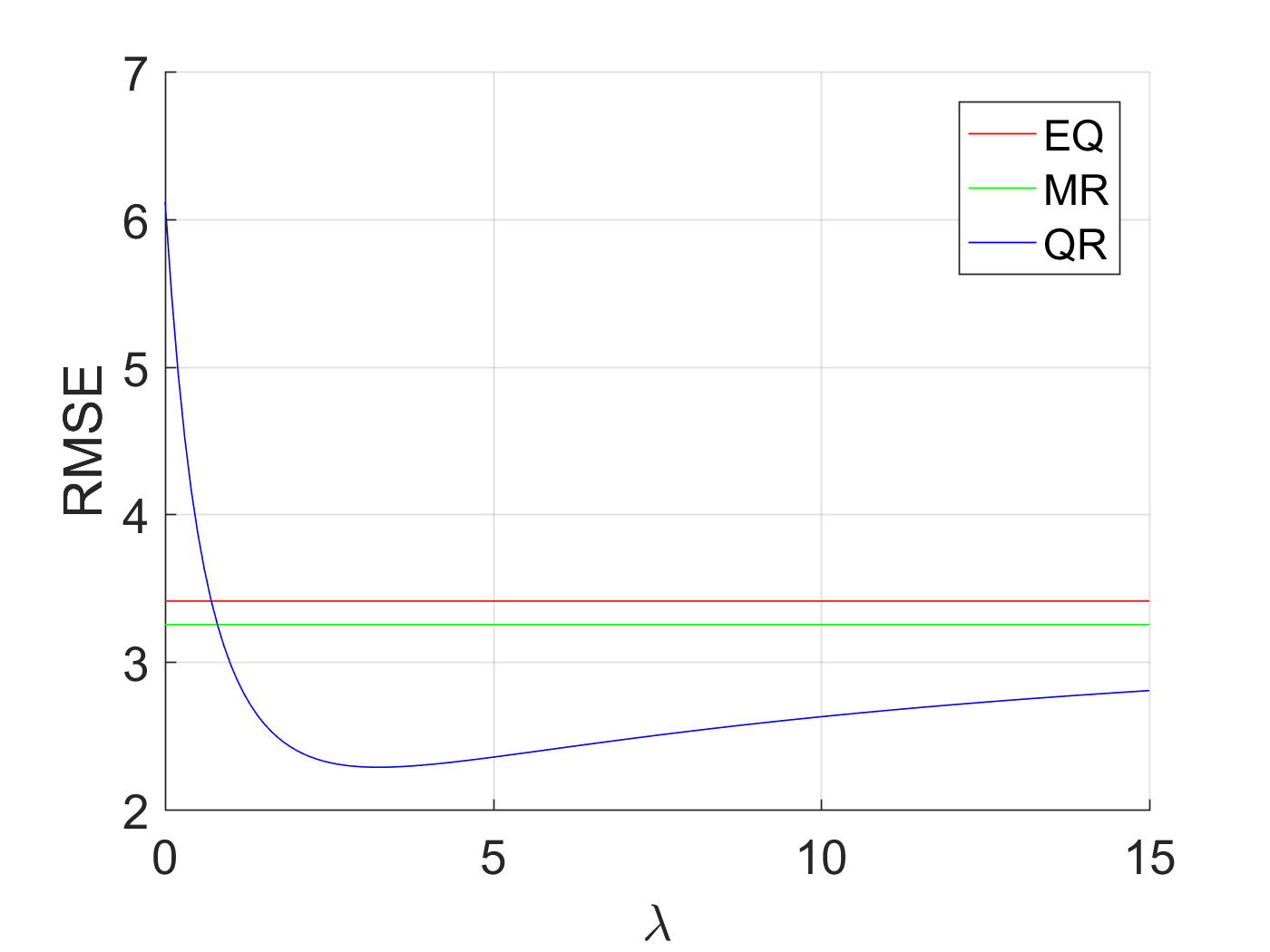}
  \caption{In 2x2 games. \label{fig:lambda-games}}
\end{subfigure}
\begin{subfigure}{.32\textwidth}
  \includegraphics[scale=0.13]{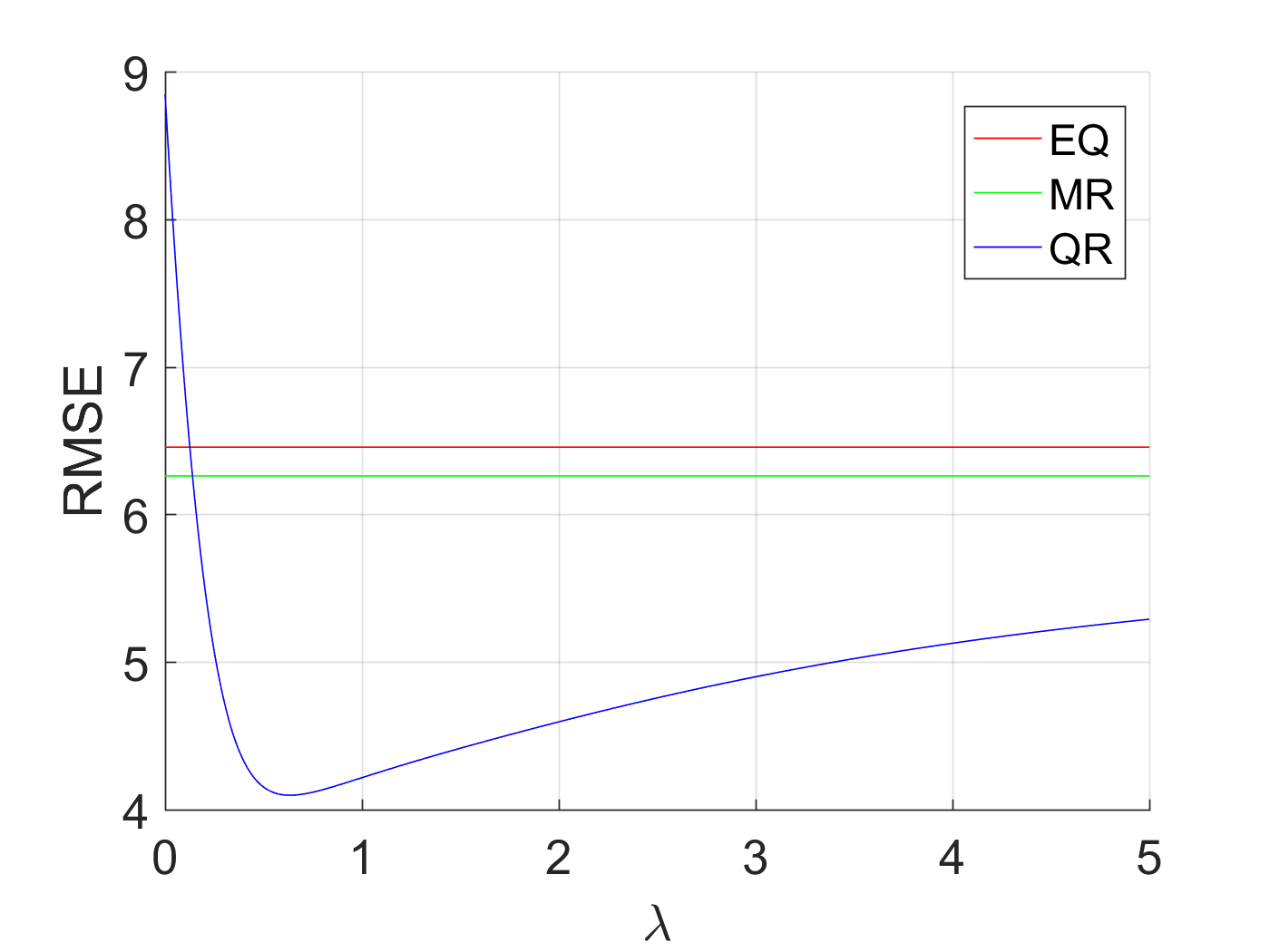}
  \caption{In VCG auctions. \label{fig:lambda-vcg}}
\end{subfigure}
\begin{subfigure}{.32\textwidth}
  \includegraphics[scale=0.13]{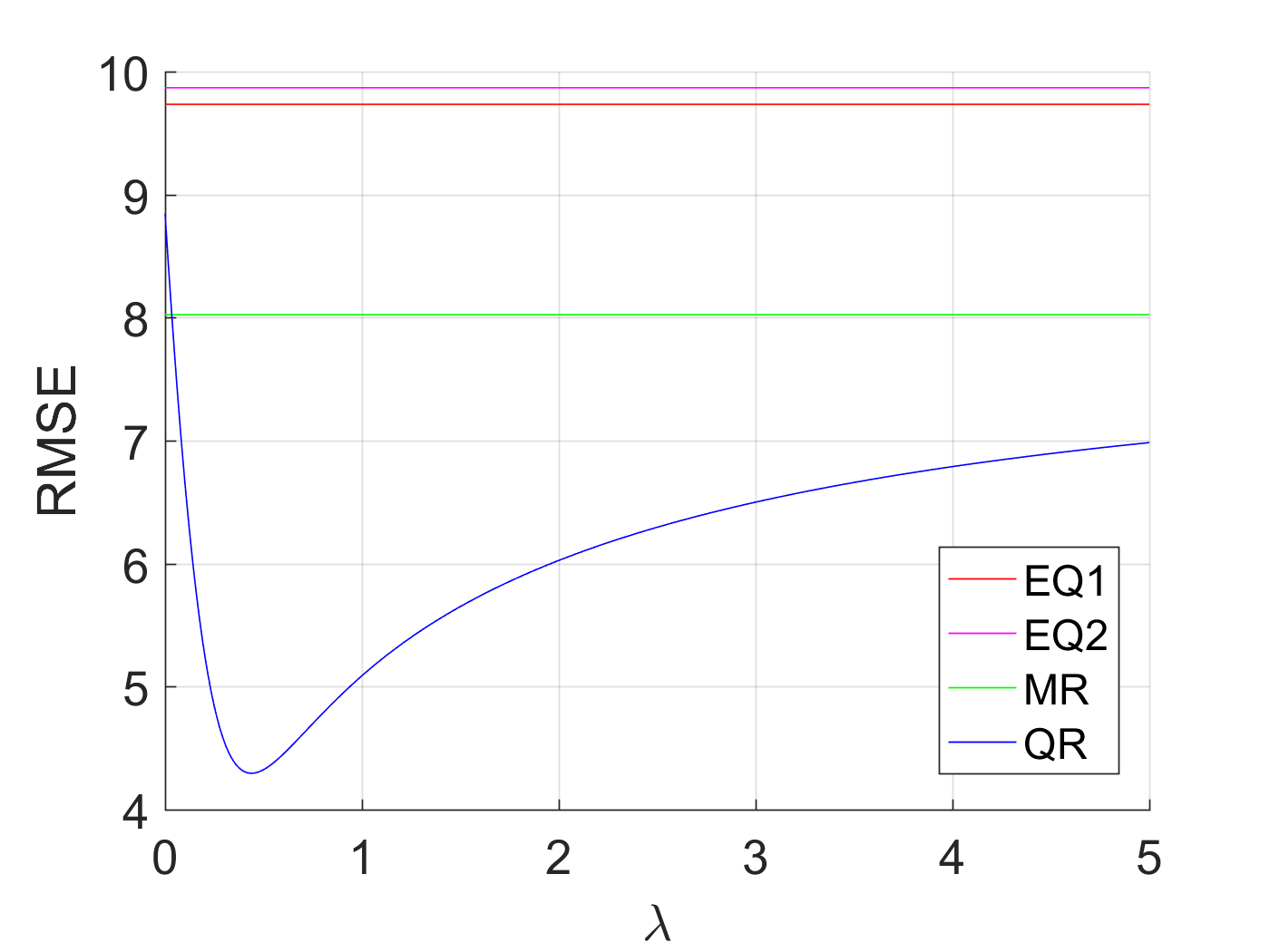}
  \caption{In GSP auctions \label{fig:lambda-gsp}}
\end{subfigure}
\caption{The RMSE of the quantal regret method using different values of the regret aversion parameter $\lambda$, compared with the RMSE of the min-regret method and the classic equilibrium-based method, for 
all sessions of the 2x2 game dataset (\ref{fig:lambda-games}), and for the VCG and the GSP sessions of the ad auction dataset (\ref{fig:lambda-vcg} and \ref{fig:lambda-gsp}, respectively) .  \label{fig:lambda}}
\end{figure}

Moreover, these results were very robust: First, the quntal-regret method outperformed the two other methods for
each one of the 12 games studied (except for a single one where all three methods gave nearly identical excellent results).  
Second, we tried
several variants of handling the nuanced details of the experiment, and all gave similar results.  Third, the method
was robust to varying
the range of valuations or the grid size that were used for the regret calculation.  Finally, the method is  robust
to the choice of the regret aversion parameter $\lambda$, and
Figure \ref{fig:lambda-games} shows that the quantal regret outperforms the other two methods 
for a very wide range of values. 
See Section \ref{sec:games} for more details regarding the setup and the analysis of the 2x2 game dataset.  

%\subsubsection{Ad Auctions}
Our second dataset is from an ad-auction experiment that we ran \cite{NNY2014}.  Each session in this experiment had 5 ``advertisers'' repeatedly competing for 5 ad-slots, with differing ``click through rates''
in a sequence of 1500 auctions.
This experiment used both the common ``GSP'' auction rule
and the theoretically appealing ``VCG'' auction rule, and also varied 
whether players were explicitly given their values or had to ``learn'' them from their feedback.   
For each setting there were 6 different sessions (all together 24 sessions with a total of 120 bidders).
Each of the five bidders in a session was given a value per click from a fixed set of values.  The
econometric task tries to recover the value of each of the players only from the sequence of bids of all players in the repeated auction.  
As mentioned above, in a previous paper \cite{NN2017} we compared classical equilibrium-based
methods to the min-regret method suggested by \cite{Nekipelov2015}, and the bottom line is that there
was a slight -- usually not statistically significant -- advantage to the min-regret method.  
Here we find, as summarized in Table \ref{fig:bottom-auctions}, that the qunatal regret method has a large and 
statistically significant advantage over both.  

Again, the advantage of the quantal regret method is very robust: it holds for each of the %four experimental conditions
different experimental settings 
(GSP vs. VCG and ``Given-Value'' vs. ``Deduced-Value'') separately, as well as %for each of the player types.  
in almost all of the comparisons according to player types. 
It is also 
robust to variants of implementation used in the previous paper, such as discarding an initial learning phase or 
looking at relative error rather than absolute one.  Finally, as demonstrated in Figures \ref{fig:lambda-vcg}
and \ref{fig:lambda-gsp}, the results hold for a wide range of values of the regret aversion parameter $\lambda$.
We provide a more detailed description of the setting as well as describe our analysis in Section \ref{sec:auctions}.

%\section*{}
The rest of the paper describes and discusses the qunatal regret method in detail in Section \ref{sec:method}, and then goes into the specific analysis  of our two datasets in Sections \ref{sec:games} and \ref{sec:auctions}.  We believe that the quantal regret method is applicable and can give 
%relatively precise estimates in
estimates that are much more precise than existing methods in 
a variety of repeated game settings.  %Obviously 
The validation or refutation of this on further datasets is the main challenge that we leave.

\section{The Quantal Regret Method} \label{sec:method}

% method.tex

\subsection{Setting}

We are looking at $n$ players repeatedly playing a game.  
We have partial information about the game, where the unknown information is captured by a parameter
$\theta \in \Theta \subseteq \Re^m$, and the game is defined by known utility functions $u_i(a_i, a_{-i}, \theta)$.
We are given the actual play of the players for a duration of $T$ repetitions of the game: at each time step $t=1...T$,
each player $i$ has played the action $a_i^t \in A_i$, where $A_i$ is player $i$'s action space.  Our task is to estimate
the unknown parameter $\theta$ given the observed behavior $\vec{a}=((a_1^1,...,a_n^1),(a_1^2,...,a_n^2),...,(a_1^T,...,a_n^T))$ of the players.  
We also have some prior $p(\theta)$ on the possible values of $\theta$.  

In the most common special case we have that (1) $\theta=(\theta_1,...,\theta_n)$, where $\theta_i$ 
captures all unknown information
about the ``value'' of player $i$, i.e., that the utility functions are given by $u_i(a_i, a_{-i}, \theta_i)$; 
(2) the $\theta_i$'s are independent of each other, i.e., 
$p(\theta)=p_1(\theta_1) \cdot \cdots \cdot p_n(\theta_n)$; and (3) the distribution $p$ is simply uniform over $\Theta$ 
(and thus also the marginal distributions $p_i$ are uniform).

\subsection{The Method} \label{subsec:method}

The quantal regret method takes the actual play $\vec{a}$ as input and 
computes (in principle) for each possible $\theta$
and for each player $i$, the player's {\em regret} had his utility in the game been defined by this value of $\theta$:
$$regret_i(\theta,\vec{a}) = \frac{1}{T}(max_{a'_i \in A_i} \sum_{t=1}^T u_i(a'_i,a_{-i}^t,\theta)
 - \sum_{t=1}^T u_i(a_i^t,a_{-i}^t,\theta)).$$
Our prediction from player $i$ is given by the weighted average of the possible values of $\theta$, 
where the weight of
each $\theta$ is updated from the prior in a way that is exponentially decreasing with its total regret.   I.e.:
\begin{equation} \label{eq:theta-est}
 \hat{\theta} = Z^{-1} \cdot \sum_{\theta} p(\theta) \cdot e^{-\lambda \cdot \sum_i regret_i(\theta, \vec{a})} \cdot \theta,
\end{equation}
where the normalization constant $Z$ is given by  
$Z=\sum_{\theta} p(\theta) \cdot e^{-\lambda \cdot \sum_i regret_i(\theta, \vec{a})}$.

In the common special case defined above, the regret of each player $i$ only depends on $\theta_i$, which implies that
each $\theta_i$ can be evaluated separately:
$ \hat{\theta_i} = Z^{-1}_i \cdot \sum_{\theta_i} \cdot e^{-\lambda \cdot regret_i(\theta_i, \vec{a})} \cdot \theta_i $,
where the normalization constant is given by  
$Z_i=\sum_{\theta_i} \cdot e^{-\lambda \cdot regret_i(\theta_i, \vec{a})}$.
In natural cases where the $\theta_i$ are scalar (or low-dimensional) it is %easily 
feasible to literally 
compute $regret_i(\theta_i, \vec{a})$ for every possible value 
of $\theta_i$ (at least in a sufficiently fine grid), and then explicitly compute the weighted expected value with these weights.  

We note that this method is %very 
general and quite easy to implement: calculating the regret of a given player for a given
possible value of $\theta$ and an empirical sequence of play $\vec{a}$ is straightforward from the definition, as long as the
set of possible actions of the player in the stage game is small enough.  For a continuous range of actions, taking a fine enough grid 
will certainly suffice.  The method requires also going over all possible values of $\theta$. In the %basic 
common case where 
$\theta=(\theta_1,...,\theta_n)$ and each $\theta_i$ captures the information for a single player $i$, as mentioned
above, each of these $\theta_i$
can be computed separately, and each of these is typically low-dimensional enough as to
feasibly allow trying all values in a reasonably fine grid.

This method is parametrized by the {\em regret aversion} parameter $\lambda$.
A value of $\lambda$ close to zero corresponds to 
the regret having no significant effect on players' behavior, in which 
case we get very little information from the empirical play
and our estimate is simply the expected value of $\theta$ according to the prior distribution.  
A very large value of $\lambda$
implies that players are much more likely to act in a way that has less 
regret than one with more regret, and thus
the prediction becomes the maximum likelihood $\hat{\theta}= argmin_{\theta} regret(\theta)$, which is the
basic form of the suggestion in \cite{Nekipelov2015}.  
A reasonable choice for the value for $\lambda$ would depend on our assesment
of the minimal gap in utilities $\Delta$ that our human players would consider 
significant in %this type of game.  
the specific type of game that is considered. A reasonable rule of thumb for games with non-negative ``values''
could take $\Delta$ to be a few percentage points of a
player's typical value
in the game. 
Choosing $\lambda$ to be the inverse of $\Delta$ would mean that we assume that 
a regret that is larger by $\Delta$ amount is $e$ times less likely.  
For high-stakes games, or
games that are well understood by the players, 
or if the prior knowledge of the range is especially poor, 
$\lambda$ could reasonably be somewhat increased. 
In this paper we use ``round'' values of $\lambda$ that correspond to $\Delta$ being
about $3\%$ of the ``average value'' in the game.  We also discuss the empirically optimal values
of $\lambda$ and demonstrate that rather wide ranges of values 
of $\lambda$ still give good estimates.

%\subsection{Example: first price auction}
\subsection{The Procedure} \label{sec:procedure}

Next we describe how to apply the quantal regret estimation method. 
%For concreteness, here is a simple example on how to apply the quantal regret method in a first price auction setting.  
For concreteness, we exemplify all the technical details in a simple first-price auction setting, and the generalization to any other game setting is straightforward. 
In our first-price auction, 
each player $i$ has a private value $\theta_i$ for the auctioned item, which we do not know and seek to estimate.  We may 
assume that we do know some bounded range $[0,M]$ from which these $\theta_i$ may come, and
due to any lack of further knowledge we assume a uniform prior over this range.  We observe 
a sequence of bids of the players $(a_1^t...a_n^t)$  for $t=1...T$ repetitions of this auction (each time for a fresh copy of the good, where players'
values remain constant throughout). %, and it is assumed that these values also do not depend on any previous wins).  
%How can we estimate the $\theta_i$'s?

An equilibrium-based approach would perhaps assume that players somehow approach a 
Nash equilibrium of the full information game (since
the repetitions presumably allowed them to learn %any 
the 
information required for that), 
which has the highest-value player bidding just above the second-highest value,
and the second-highest player bidding his true value. The actual value of the highest 
bidder and the values of other bidders do not affect this
equilibrium and in principle could not be estimated using the equilibrium assumption.

Here, in contrast, is an implementation of the quantal regret method:

\begin{enumerate}
\item 
First we choose a reasonable grid on $[0,M]$ of possible
values.  Since it is hard to believe that our estimate will have less than a $1\%$ error
(of the interval size), it seems that taking the grid points $\{M \cdot j / 100 \:|\:j=0...100\}$ is good 
enough.  Let us denote the points on the grid by $\theta^j = j \cdot M / 100$.
\item
For notational convenience, let us denote by $W_i=W_i(\vec{a})$ the set of time steps in which $i$ won the auction, i.e., $\{t \in 1...T\:|\:a_i \ge a_k \:for\:all\:k \ne i\}$.\footnote{
Throughout this example we assume that $i$ wins ties; more generally, one should adjust the expressions according to the tie breaking rules. 
}
\item
Next, for each player $i$ and each possible
grid value $\theta^j$, we calculate the regret of player $i$ had his value been $\theta^j$:

\begin{enumerate}
\item 
The player's empirical utility had his value been $\theta^j$ is 
$u_i(\theta^j, \vec{a}) = \theta^j \cdot |W_i| - \sum_{t \in W_i} a_i^t$.

\item To calculate the optimal fixed bid $b^*_i$ for player $i$, had his value been $\theta^j$, let
$w_i(b) = |\{t | b \ge a_k\: for\:all\:k \ne i\}|$
be the
number of times that bidding $b$ would make player $i$ the winner of the auction.  So 
$b^*_i = argmax_b  w_i(b) \cdot (\theta^j -b)$.  Algorithmically, one may perform this 
maximization by exhaustive search over the possible $b$. In principle, one has to go over all
maximal bids of other players, but often there will be some bidding grid that is implied by the auction
setting, 
or one may alternatively use our $\{\theta^j\}$ grid for the bids. %too.

\item Now the regret is: 
$r_i^j = regret_i(\theta^j, \vec{a}) = w_i(b^*_i)\cdot (\theta^j - b^*_i) - u_i(\theta^j, \vec{a})$
\end{enumerate}

\item
We now need to choose a value for $\lambda$.  One may roughly follow the 
rule of thumb suggested above and take $\lambda=1/\Delta$, where
$\Delta$ is, say, $3\%$ of the average bid. %(this takes into account that the bids are somewhat smaller than the values of the bidders.)

\item 
Our total estimate for each $\theta_i$ would be 
$\hat{\theta_i} = (\sum_j \theta^j \cdot e^{-\lambda r_i^j}) / (\sum_j e^{-\lambda r_i^j})$.
\end{enumerate}

\subsection{Rationale}

The basic assumption behind this method is that players tend to succeed in minimizing their regret in a long repeated play, 
%although 
however 
they do not do so perfectly but rather are less likely to achieve higher regret than to achieve lower regret.
This assumption -- which this paper tests empirically -- makes sense from several perspectives.  

As argued in \cite{Nekipelov2015}, the basic ability
of players to learn enough about the environment as to be able to minimize their regret is a rather weak assumption, 
and in particular is strictly weaker
than assuming that they reach a Nash equilibrium or even a correlated \cite{Corr1974} or coarse-correlated \cite{Coarse} one (since in
any such equilibrium the players must all be minimizing their regret). %\cite{}).  
Furthermore, there are many natural dynamics
that are known to indeed minimize regret in the long run -- see, e.g., \cite{BM2007, Arora2012, Hart2000}.  \\ Note also that such dynamics
are possible even when players have very little knowledge about the game that they are playing, but only
receive feedback about their utilities (the ``bandit'' setting). 
Nevertheless, it is worth emphasizing that this assumption is far from trivial in two senses: first that the players 
are sufficiently smart and rational and posses sufficient information as to succeed in minimizing their regret, and,
second, that the players are not even smarter and cooperative as to reach a threat-based equilibrium of the repeated game
which may have high regret even though it may Pareto-dominate all equilibria of the stage game.

It is possible to justify the quantal regret method as the one that minimizes the expected square error after a
Bayesian update of the prior  based on the assumption that
the regret affects the probability that a player plays any particular
sequence of actions $(a_i^1...a_i^T)$ in a way that is proportional to $e^{-\lambda \cdot regret_i(\theta, \vec{a})}$.
Formally our assumption is that 
$Pr[\vec{a}|\theta] = z^{-1} \cdot Pr[\vec{a}] \cdot e^{-\lambda \cdot \sum_i regret_i(\theta, \vec{a})}$ where
$Pr[\vec{a}]$ is some prior likelihood of the sequence $\vec{a}$ and $z$ is just a 
normalization constant ensuring that all probabilities sum to $1$:
$z = \sum_{\vec{a}} Pr[\vec{a}] \cdot e^{-\lambda \cdot \sum_i regret_i(\theta, \vec{a})}$.
While in this formula $z$ may depend on $\theta$, we further assume, simply due to lack of any knowledge,
that it is a constant.\footnote{This assumption cannot really be interpreted as a direct reasonable model of the
world, but since our assumption regarding $Pr[\vec{a}|\theta]$ is really quite under-specified (i.e. it could be that each
$\theta$ has a separate set of possible $\vec{a}$ in it support), together these assumptions suggest our lack of 
knowledge about $\theta$ beyond that given by the prior.}

Once we take this behavioral assumption then our estimate is simply the one that minimizes the posterior 
expected square distance
between $\hat{\theta}$ and $\theta$.    Specifically, using Bayes' rule on our assumption that
$Pr[\vec{a}|\theta] = z^{-1} \cdot Pr[\vec{a}] \cdot e^{-\lambda \cdot \sum_i regret_i(\theta, \vec{a})}$ yields that
the posterior probability over $\theta$ after observing the sequence of actions $\vec{a}$ to be:
$Pr[\theta|\vec{a}]= p(\theta) \cdot z^{-1} \cdot e^{-\lambda \cdot \sum_i regret_i(\theta, \vec{a})}$. 
The estimate $\hat{\theta}$ that minimizes
the expected square error, 
is exactly the weighted average according to this posterior, which is how we chose our estimate.

\section{Estimation in 2x2 Games} \label{sec:games}
%games.tex

\subsection{The 2x2 Game Dataset} \label{sec:games-exp}

The first dataset that we use to evaluate the quantal regret method is from an experiment of repeated 2x2 games of \cite{Selten2008}.
In their experiment they 
investigated 12 games for two players (``row'' and ``column'' players), 6 of which were constant sum and 6 non-constant sum games. 
Each of the games had non-negative utilities and was ``completely mixed'', i.e., had only one equilibrium point in which every pure strategy is used with positive probability. The utility matrices of the games are fully depicted in Appendix \ref{app:games}.  
There were 12 independent subject groups (``sessions'') for each constant sum game and 6 for each non-constant sum game, to a total of 108 sessions. Each session consisted of eight human players -- four in the role of the row players and four in the role of the column players.\footnote{In fact, \cite{Selten2008} ran two independent subject groups of eight players each in every session, however here we refer to each independent subject group as a separate session. For the full details regarding the experimental setup see \cite{Selten2008}.} 

In the beginning of every session, the players were informed about the game matrix, including the payoffs of both players. 
They interacted over 200 periods, always in the same role, and were re-matched in every period within their subject group. %Each player remained in the same role in all periods. 
In each period every row player had to choose between two actions $Up$ and $Down$, and every column player chose between $Left$ and $Right$. 
After each period, every player received feedback about the other player's choice and payoff, period number, and their cumulative payoff. 
In the end of the experimental session, every player was paid proportionally to his accumulated payoff in addition to a show up fee. 

The data consist of the utility matrices of the 12 games, as well as of the empirical frequency of play 
for each player over his 200 plays. 
Specifically, for each of the 8 players, in each of the 108 sessions, 
we observe the empirical frequency of each of the 4 strategy profiles: 
$(Up, Left)$, $(Up, Right)$, $(Down, Left)$, $(Down, Right)$, during the 200 periods. 
For illustration, Figure \ref{fig:example-mtx-x-y} presents one of the games in the dataset (with one pair of utilities hidden for estimation), and Figure \ref{fig:example-freq} shows the average of the empirical frequency of play that was obtained in one of the sessions of this game.

\subsection{The Quantal Regret Method in the 2x2 Game Setting} \label{sec:games-results}

Our econometric estimation task in the 2x2 game setting is to estimate each of the 8 parameters defining the game (the 4 utilities of the row player and the 4 of the column player), using the observed data. 
Specifically, we ``hid from ourselves'' each of the 8 parameters, one at a time, and attempt to infer this hidden parameter using the empirical frequency of play and the rest of the (unhidden) parameters. 

In this section we define the estimation in the {\em session level}, which is the general level that ignores the specific experimental setup described above, 
and in the next section we show the robustness of the results to other %possible definitions. 
levels of estimation. %including one version that considers the exact details of the experiment.  
In the session-level approach, we compute for each independent subject group (session) the empirical play by averaging the empirical frequencies of the eight players in the group (for each of the four strategy profiles). 
We consider this average %(which of course also sums to 100\% over the four profiles) 
as if it was the empirical frequency played by two (global) players in the session, and use it to estimate the 8 parameters in the game  -- 4 parameters of the (global) row player and 4 of the (global) column player.  %Thus, for each of the 108 sessions (groups) we estimate (separately) each of the 8 parameters defining the game. 

We applied the quantal regret method according to the procedure described in Section \ref{sec:procedure}, to estimate the 8 parameters in each session. 
In this section we set the regret aversion parameter by $\lambda=3$, which corresponds to $\Delta$ of about 3\% of the average value in the games (as suggested in Section \ref{subsec:method}), and 
for each session, the regret for each of the two players was computed for each of his 4 parameters over the integers in the interval $[0,22]$. %The regret aversion parameter $\lambda$ was set to $3$ in the basic implementation, and 
In the next section we show the robustness of the results to different implementation variants.  

We compare the estimates obtained using the quantal regret (QR) method with those obtained using the min-regret (MR)  
and the equilibrium-based (EQ) methods. As described in Section \ref{sec:intro}, the min-regret method takes as the estimate the value that minimizes the regret. The equilibrium-based method assumes that the empirical frequency of play is the mixed Nash equilibrium of the game, and utilizes it to derive value estimate for each of the parameters, as was demonstrated in Section \ref{sec:intro}. For a fair comparison, since the estimates of the regret-based methods are restricted to the parameter range $[0,22]$, 
we keep the EQ estimates in the same range. %valuation range used for the regret based methods ($[0,22]$). 
We define the session estimation error as the RMSE over the estimation errors of the 8 parameters in the session, and, as explained in Section \ref{sec:intro}, we compare the success of the estimation methods based on the RMSE, on the average of the estimation errors, and on the $\pm 3$ hit-rate, in a given setting.

Table \ref{fig:bottom-games} presents the bottom line of the comparison results over all 108 sessions in the experiment. 
It clearly shows that the quantal regret method outperforms the two other methods, which in turn have similar performance with a small advantage to the min-regret method (as could be expected, since the MR method rounds the EQ estimates in the correct direction, see Section \ref{sec:intro}). Specifically, the RMSE of the QR method was far lower than the RMSE of the MR and the EQ methods, and the difference was statistically significant: the estimation errors using the QR method were significantly lower than the errors using each of the other two methods (paired two-sided Wilcoxon signed rank test, N=108 sessions, $p<0.0001$). In addition, although smaller, the difference between the errors of the MR and the EQ methods was statistically significant ($p<0.0001$).
Moreover, Table \ref{fig:bottom-games} shows that the quantal regret method has the highest hit-rate, with 81.6\% of the parameter estimates within a delta of 3 from the true values. %, in comparison with hit-rates of 75.0\% and 68.87\% using the MR and the EQ methods, respectively. 

The quantal regret method outperforms the other two methods also when considering the constant and the non-constant sum games separately. 
Table \ref{tbl:rmse-cs-vs-ncs} shows the estimation results over the 72 sessions of the constant sum games and over the 36 sessions of the non-constant sum games. 
It can be seen that for both types of games, the RMSE of the QR method is lower than the RMSE of the other methods, and the differences were statistically significant at the 1\% level in both cases (paired two-sided Wilcoxon signed rank tests, N=72 or 36 sessions). 
Also the errors using the MR method were significantly lower than those using the EQ method at the 1\% level for both types of games. 
Moreover, Table \ref{tbl:rmse-12-games} shows that the quantal regret outperforms the other methods in each of the 12 games separately; except for Game 6 for which all methods perform very well, the QR method has lower RMSE than the others in estimating each of the games, and usually by a large gap.

\begin{table}
\centering
\begin{tabular}{|l||l l l||l l l|}
\hline
& \multicolumn{3}{|c||}{Constant Sum Games} &  \multicolumn{3}{|l|}{Non-Constant Sum Games} \\ \hline
&   EQ & MR & QR &  EQ & MR & QR \\ \hline
RMSE &  3.42 & 3.27 & 2.14 &  3.39 & 3.23 & 2.57   \\ 
Avg Err & 3.02 & 2.86 & 1.93  &   2.93 & 2.81 & 2.25  \\
$\pm 3$ Hit Rate & 69.27\% & 75.00\% & 83.68\% &   68.06\% & 75.00\% & 77.43\% \\
\hline 
\end{tabular}
\caption{Estimation results in the 2x2 game dataset, 
for the constant-sum sessions (on the left) and for the non-constant sum sessions (on the right). The full details of the game-level results are provided in Table \ref{tbl:rmse-12-games}. \label{tbl:rmse-cs-vs-ncs}}
\end{table}

\subsection{Robustness of the Results} \label{sec:games-robust}

After seeing that one implementation of the quantal regret method manages to estimate the 2x2 game dataset better than the other two methods, 
let us now show that this is robust to other implementation variants. 

\subsubsection*{Robustness to the estimation level}

In the previous section we took a general approach and tested the session-level estimates. 
We tried several variants of handling the specific setup of the experiment, 
where in every session there were 4 row players and 4 column players, who were re-matched in every period, and thus interacted directly or indirectly with each other (see Section \ref{sec:games-exp}). 
First, we estimated the parameters in each session by a {\em fine grained aggregation} of the players, i.e., by aggregating the 4 row players and the 4 column players separately. For the regret-based methods QR and MR, this implies deriving the 4 parameters of players of the same type by their {\em total} regret. For the QR method this is exactly like Equation \ref{eq:theta-est} specifies, with $\lambda$ reduced proportionally to 3/4. %(as the regrets are summed over 4 players).  
Table \ref{tbl:levels} shows that although this fine grained aggregation of the players improved the MR results relative to its session-level results, it hardly affected the QR results, and overall the results remained qualitatively the same. 

Second, we tried to derive the parameter estimates %from {\em each player separately}, 
in the {\em player-level}, 
obtaining 4 estimates from each of the 8 players separately for each session. This approach, that derives the estimates based on less information and is thus more prone to noise, increased the error for all three methods, as could be expected. 
Third, taking a more general approach and deriving the parameter estimates in the {\em game-level} (i.e., based on the average of the empirical frequencies over all sessions for a game), improved the estimation results for all three methods, as expected. 
Table \ref{tbl:levels} summarizes the estimation results and shows that overall the results reported in the previous section, where the QR method outperforms the other methods, are robust to the different estimation levels.

\begin{table}
\centering
\resizebox{\textwidth}{!}{ %scale down table to the textwidth
\begin{tabular}{|l|l|l l l|}
\hline
Estimation Level & & EQ & MR & QR  \\ \hline
Game-level estimate  & RMSE & 3.12 & 2.91 & 2.01  \\ 
& Average Error  & 2.71 & 2.54 & 1.76 \\
& $\pm 3$ Hit Rate & 72.92\% & 77.08\% & 82.29\% \\
\hline
Session-level estimate (Baseline) & RMSE  & 3.41 & 3.25 & 2.29 \\ 
& Average Error  & 2.99 & 2.84 & 2.04 \\
& $\pm 3$ Hit Rate & 68.87\% & 75.00\% & 81.60\% \\
\hline
Fine grained aggregation of players  & RMSE & 3.41 & 3.04 & 2.30 \\ 
& Average Error  & 2.99 & 2.63 & 2.05 \\
& $\pm 3$ Hit Rate & 68.87\% & 78.70\% & 81.13\% \\
\hline
%Separate estimate from each player & RMSE & 4.72 & 3.40 & 2.95 \\ 
Player-level estimate & RMSE & 4.72 & 3.40 & 2.95 \\ 
& Average Error & 4.57 & 3.08 & 2.77 \\
& $\pm 3$ Hit Rate   & 56.66\% & 73.15\% & 73.29\% \\
\hline
\end{tabular}
}
\caption{Estimation results over all 108 sessions of the 2x2 game dataset, for different estimation levels. 
\label{tbl:levels}}
\end{table}

Lastly, 
while up to here the estimation was for a general 2x2 game, 
estimation for the constant sum games can take into account the constant sum property of the game. 
In this case, the estimation task is of estimating only 4 parameters rather than 8 (e.g., the 4 payoffs of the row player), and these parameters can be estimated from the aggregation of the two players in the session-level. 
In Appendix \ref{app:games-robust} we show that, as expected, 
the aggregation of the two players improves the estimation results 
for all the three estimation methods. We note though that the improvement for the MR method is sharper and in the aggregated case it outperforms the QR method.

\subsubsection*{Robustness to the parameter selection}

Let us now return to the session-level estimates and show that the results presented in the previous section are robust to different selection of the parameters. 
First, while we demonstrated that the quantal regret method outperforms the other methods using the regret aversion parameter $\lambda=3$, it is in fact true for all reasonable values of $\lambda$. Figure \ref{fig:lambda-games} shows the RMSE of the quantal regret method as a function of the parameter $\lambda$, for all 108 sessions, compared with the RMSE of the min-regret and the equilibrium-based methods. 
As can be seen, the RMSE of the quantal regret method is consistently lower than the RMSE of the other methods, and as expected approaches the RMSE of the min-regret method as $\lambda$ grows large. 
The best results with RMSE of 2.29 are achieved at $\lambda=3.3$. %, and for $1.5 \leq \lambda \leq 10$ the RMSE stays below 2.6. 
For very low values of $\lambda$ where the quantal regret method effectively doesn't consider the regret levels in the estimation process, the error is  large. 

\begin{figure} % slide 12
\begin{subfigure}{.5\textwidth} 
  \includegraphics[scale=0.19]{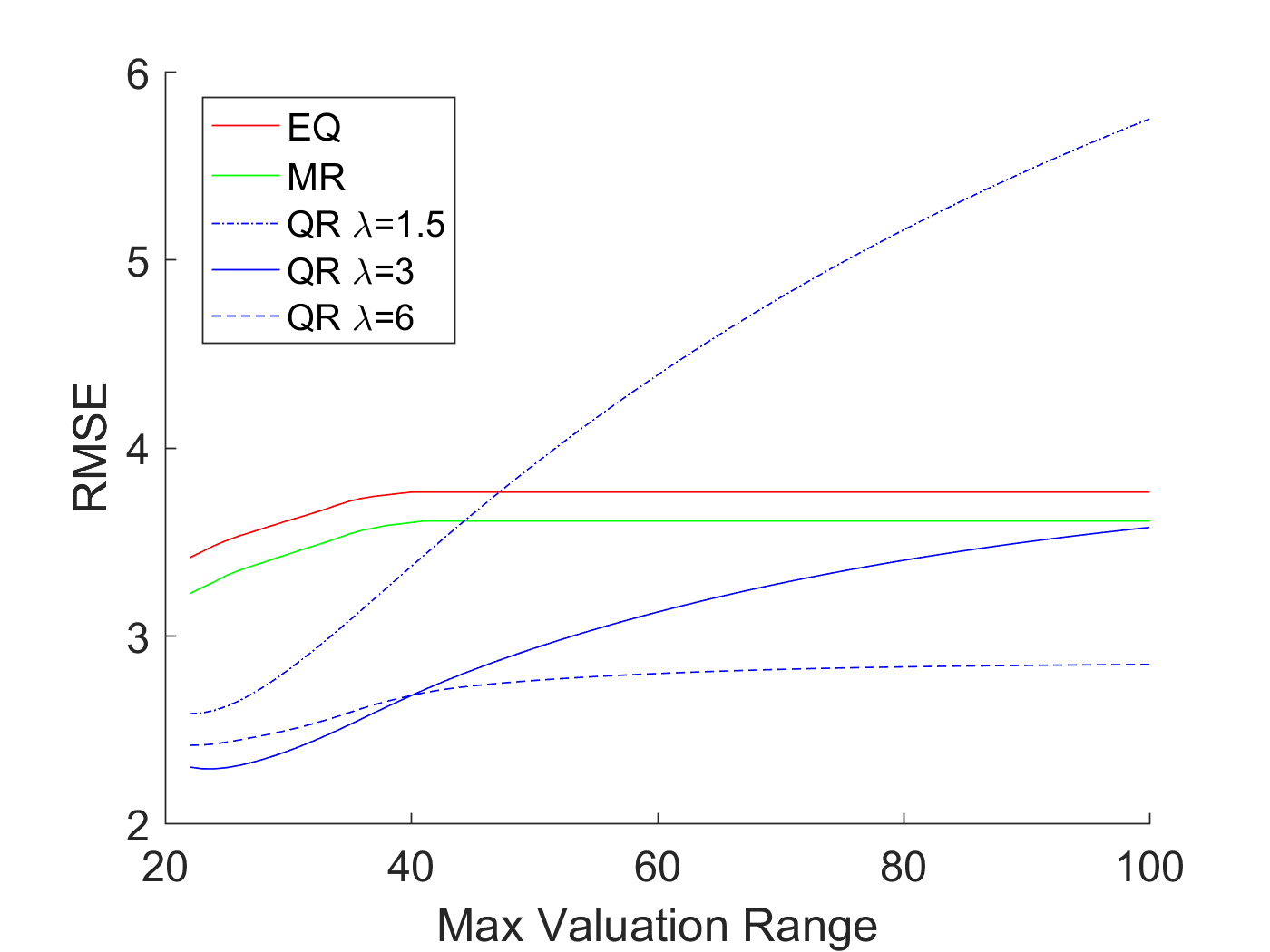}
  \caption{\label{fig:range-rmse}}
\end{subfigure}
\begin{subfigure}{.5\textwidth}
  \includegraphics[scale=0.19]{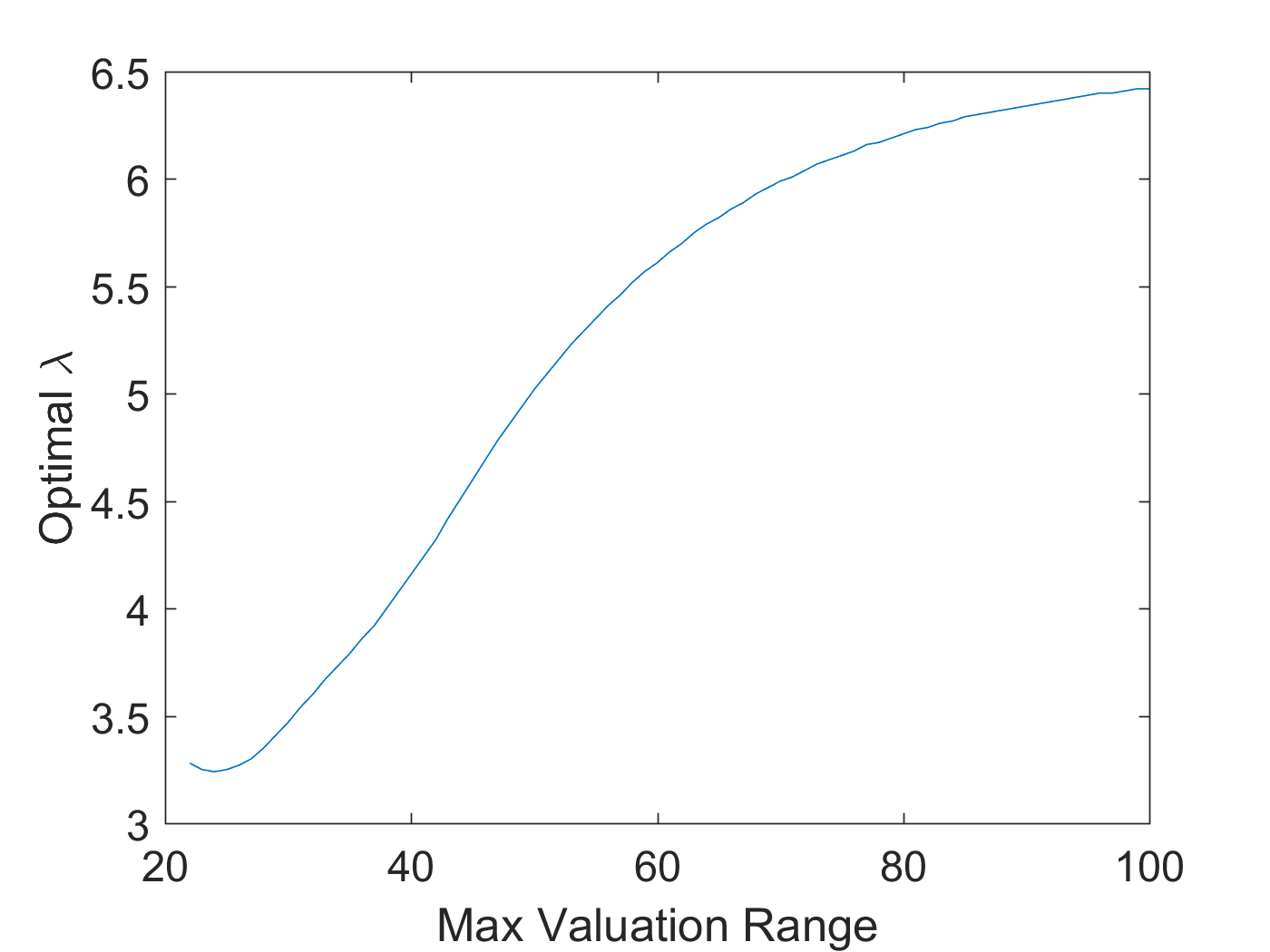}
  \caption{\label{fig:range-opt-lambda}}
\end{subfigure}
\caption{The RMSE (\ref{fig:range-rmse}) and the optimal $\lambda$ (\ref{fig:range-opt-lambda}) as a function of the upper bound on the valuation range that is considered for the estimation.  }
\end{figure}

Second, in the previous section we assumed that we know a quite accurate range of valuations $[0,22]$ from which the parameters may come, %(and assumed uniform prior over this range). 
however good results are obtained even when the prior is much less accurate, as is shown in Figure \ref{fig:range-rmse}. 
The graph also shows that with a less accurate prior it is better to use a higher $\lambda$ value, while for a good prior it is better to use a lower value, and Figure \ref{fig:range-opt-lambda} shows how the optimal value of $\lambda$ increases with the valuation range that is considered. 
Finally, Table \ref{tbl:grid} in Appendix \ref{app:games-robust} shows that the quantal regret is robust to varying the resolution of the grid being used for the regret calculation, while as expected with a more accurate grid the min-regret estimates become closer to the equilibrium-based estimates.

\section{Estimation in Ad Auctions} \label{sec:auctions}
%adauctions.tex

\subsection{The Ad Auction Dataset}
The second dataset is from a previous experiment that we ran, where 
%We performed a controlled experiment where 
human subjects were asked to participate in a simulation
of ad auctions, similar to those held by search engines like Google or Microsoft (also known as ``sponsored-search auctions'').  
This experiment was described in
\cite{NNY2014}, which contains all the details as well as the results.  
In the experiment, %we recruited participants in 
groups of five participants simulated the roles of advertisers and had to 
compete in a stream of ad auctions that lasted 25 minutes. 
%We used a flexible auction experimentation software that we developed that enabled us to control the auction details as well as the players' knowledge and values. 
The auctions were conducted continuously, one auction per 
second, 
to a total of 1500 auctions. % within the 25-minute game. 
The participants could modify their bids at any time, 
and each auction was performed with the current settings of the bids. 
Each player was assigned a ``type'' at random, %where his type 
which was
his private ``valuation,'' i.e., the monetary value that he obtained from each user 
who clicked on his ad (we used 21, 27, 33, 39, 45 ``coins'').  
Players did not know the values of the other players nor the bids that the others made. 
Each ad auction sold 
five ad positions with varying (commonly known) Click Through Rates (CTR) (we used 2\%, 11\%, 20\%, 29\%, 38\%), 
which were displayed in a decreasing order of CTRs, such that the position on the top of
the page received the highest CTR. 
Every time an advertiser with a valuation $v$ won a position 
with CTR $\alpha$, he got an income of $\alpha \cdot v$ from
that auction.  This income was added to his balance
and the appropriate payment according to the auction rule was deducted from his balance. 
The players were given a graphical user interface in which they could 
modify their bids as often as 
they wished, and follow the results of the auctions so far. 
Appendix \ref{app:screenshot} presents a screen shot of the user interface. 
In the end of the experiment players were paid for their participation proportionally 
to their final balance (in addition to a fixed participation fee). 

The experiment had a two-way (2x2) between-participant design;  
thus there were four experimental conditions. 
The two factors were:

\begin{enumerate}
\item {\bf Payment Rule (the Auction Mechanism):} The (theoretically appealing) VCG payment rule was compared with the (commonly used) GSP payment rule.
Both VCG and GSP auctions make the same allocation of positions -- by decreasing order of bids -- but their payment rule is different. Unlike GSP, the VCG is truthful; i.e., in every VCG auction it is a dominant strategy for every player to bid his true value (see \cite{EOS2007,Varian2007}).
\item {\bf Valuation Knowledge:} While the starting point of analyzing behavior in auctions is the ``valuation'' of the bidder, it is questionable to
what extent users are explicitly aware of this valuation.  We compared the case where bidders were directly given their valuation (given value, GV), 
and were explained its significance,
and the case where bidders were not directly given the valuation, but rather only see their payoffs -- information from 
which the valuation may be deduced, but could alternatively be directly used to guide the bidding (deduced value, DV).
\end{enumerate}

There were a total of 24 experimental sessions, 6 sessions for each of the
4 experimental conditions (thus there were 12 sessions for each factor). 
The groups (of five players each) were randomly assigned to the 
four experimental conditions, giving a total of $n$ = 120 participants. For further details regarding the experimental setup see \cite{NNY2014}.

\subsection{The Quantal Regret Method in the Ad Auction Setting}

Our econometric task in the ad auction setting is to estimate the private values of the bidders from the observational data in the ad auction game.
That is, we now assume that we do not know the private values and wish to recover these values from the observed sequence of bids that was played, the observed CTRs, and the utility functions of the players.  

\begin{figure*}
\begin{center}
\includegraphics[scale=0.25]{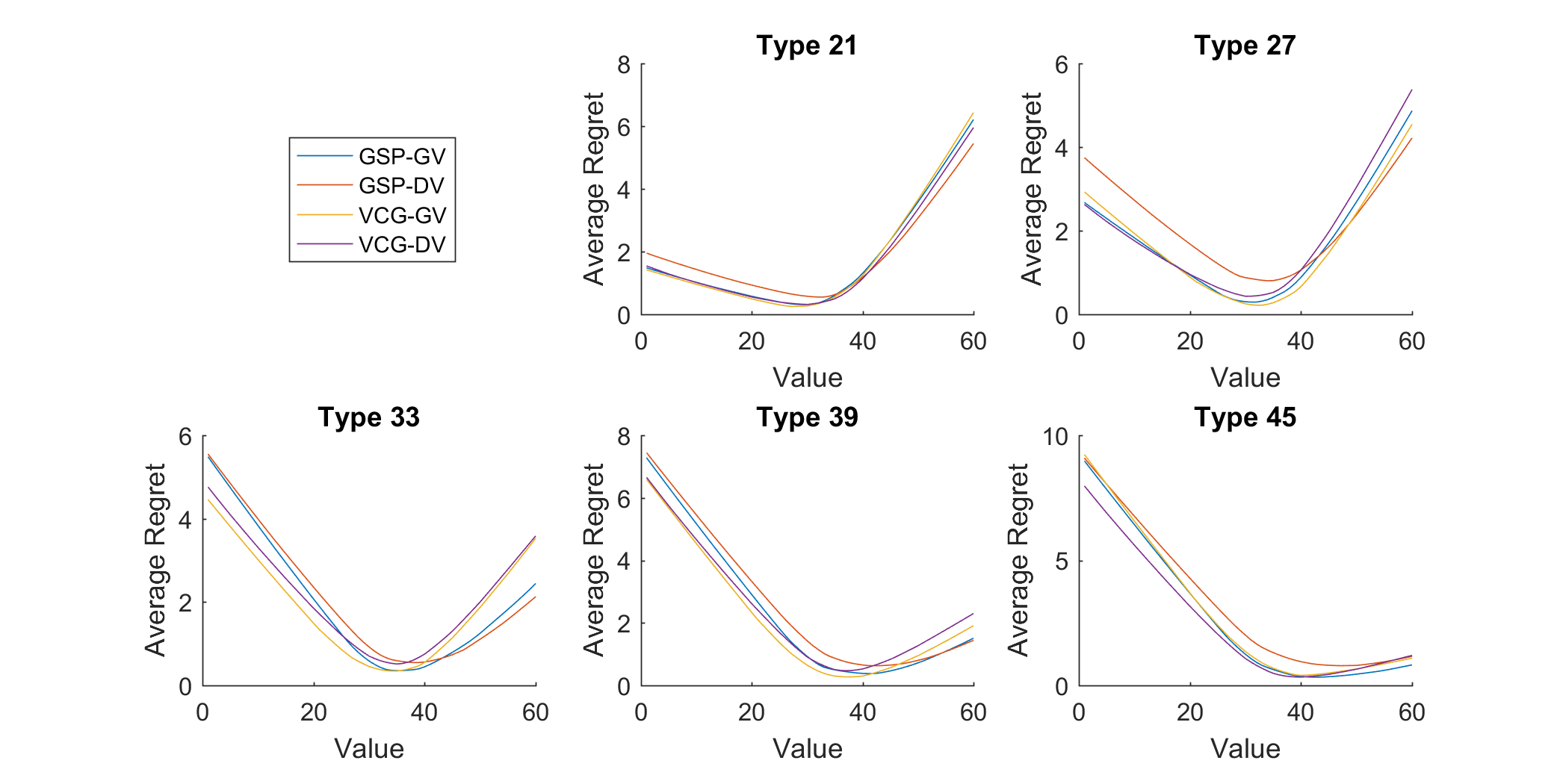}
\end{center}
\caption{The regret as a function of value for the players in the ad auction experiment, averaged according to player types and experimental conditions (computed over the 1500 auctions in the game).
 \label{fig:regret-curves}}
\end{figure*}

We applied the quantal regret method to derive value estimates for each of the players, as described in Section \ref{sec:procedure}.
We computed the regret in the ad auction game (over the 1500 auctions) for each of the bidders, over the integer values in the range $[1,60]$. Figure \ref{fig:regret-curves} presents the regret results as a function of these values, averaged over players of the same type and by the four experimental conditions. 
The min-regret method suggested by \cite{Nekipelov2015} takes the minimum point for each player as the estimate for his value, which is clearly visible in the graphs. In contrast, the quantal regret method takes a weighted average of the possible values, with weights that are exponentially decreasing with the regret. %, as described in Section \ref{sec:method}.
In this section we use $\lambda=1$ as the regret aversion parameter, which corresponds to $\Delta \approx 3\%$ of the average utility in the auction game (as suggested by the rule in Section \ref{subsec:method}), and in the next section we show the robustness of the results to different values of this parameter as well as to other implementation variants. 
 
We compare the estimates obtained using the quantal-regret method with those obtained using the min-regret method and using econometric methods that rely on the standard equilibrium assumption. 
Since in \cite{NN2017} we have already compared min-regret with the standard econometric methods on the same ad auction data, here we focus on the performance of the newly suggested quantal-regret method. 
We evaluate the results separately for GSP and VCG, since the standard methods are different for these two auctions due to their different equilibria predictions, however notice that the two regret-based methods are general and work just the same for both cases. 
%While the regret based methods are general and work just the same for both GSP and VCG auction rules, standard econometrics use different methods for these two auctions since they have different equilibria, and therefore we perform the evaluation separately, starting with the VCG auction. 
Again, our main measure for the quality of the estimation methods is the RMSE over the estimation errors in a given setting. 
Specifically, we compute for every player $i$ whose true value is $v_i$, the absolute estimation error by: $error_i=|v_i-\hat{v}_i|$, where $\hat{v}_i$ is the value estimate for player $i$. The estimation error on a set of players $S$ is the RMSE of the players in $S$: 
$error(S)=\sqrt{\frac{1}{|S|}(\sum_{i \in S}error_i^2)}$. Additionally, we look at the average estimation error and at the $\pm 6$ hit-rate achieved in a given setting, as explained in Section \ref{sec:intro}.

\subsubsection{Evaluating Estimations in VCG Auctions}

We start by considering the VCG auction which is a simple case for standard econometrics: in VCG bidding truthfully is a dominant strategy, and so it should be a strong prediction that players will all bid their true value in equilibrium. Thus, the classic econometric method is simply taking the average bid that a bidder played in a sequence of auctions as the estimate for his value in these auctions. 

Table \ref{fig:bottom-auctions} summarizes the bottom line of the estimation results over all 60 VCG bidders, for the quantal regret (QR) method, the min-regret (MR) method, and the standard equilibrium-based (EQ) method of taking the average bid. 
%and 
It 
clearly shows that the quantal regret method succeeded much better than the other two methods, which, in turn, had similar performance. 
Specifically, the RMSE of the QR method was %4.22 while the RMSE of MR and EQ were about $50\%$ higher,   
about 30\% lower than the RMSE of the others, 
and the difference was statistically significant:
the estimation errors using the QR method were significantly lower than the errors using each of the other two methods (paired two-sided Wilcoxon signed rank tests, N=12 sessions, $p<0.001$). %, with an average error of about half of that obtained using each of the other methods. 
In addition, while the QR method achieved the highest hit-rate of 81.7\%, %with 81.7\% of the estimates within a delta of 6 from the true value, 
the MR and the EQ methods had very poor hit-rates; %the EQ method, which is based on the strong prediction of the truthful dominant strategy, managed to achieve a hit-rate of only 61.7\%. 
only 61.7\%  of the estimates using the EQ method were withing a delta of 6 coins from the true value, despite the fact that this method is based on the particularly strong prediction of the truthful dominant strategy. 

\begin{figure}[t] % slide 12
\begin{subfigure}{.32\textwidth} 
  \includegraphics[scale=0.13]{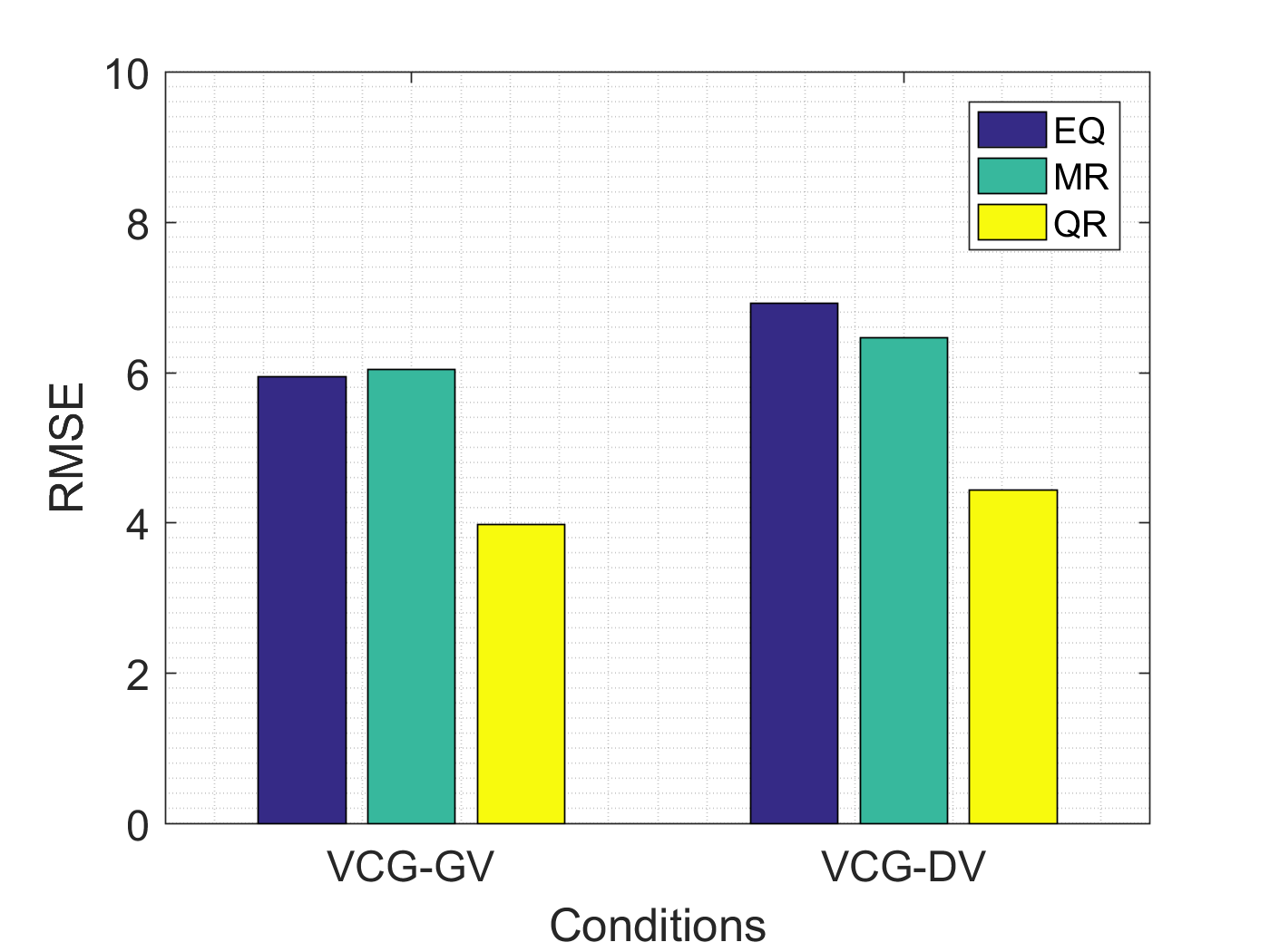}
  \caption{By information %conditions 
	\label{fig:vcg-bars-info}}
\end{subfigure}
\begin{subfigure}{.32\textwidth}
  \includegraphics[scale=0.13]{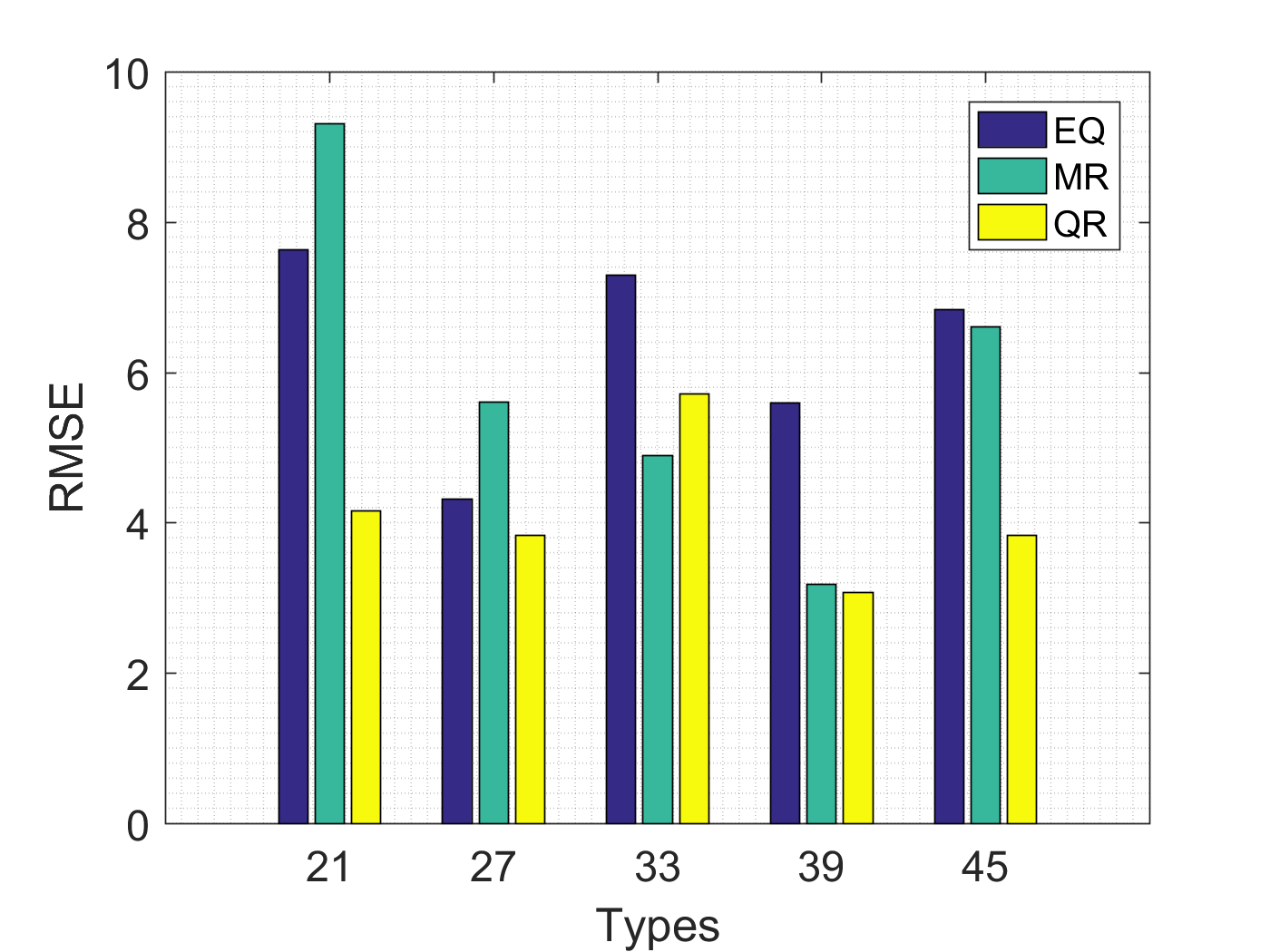}
  \caption{By player types \label{fig:vcg-bars-type}}
\end{subfigure}
\begin{subfigure}{.32\textwidth}
  \includegraphics[scale=0.13]{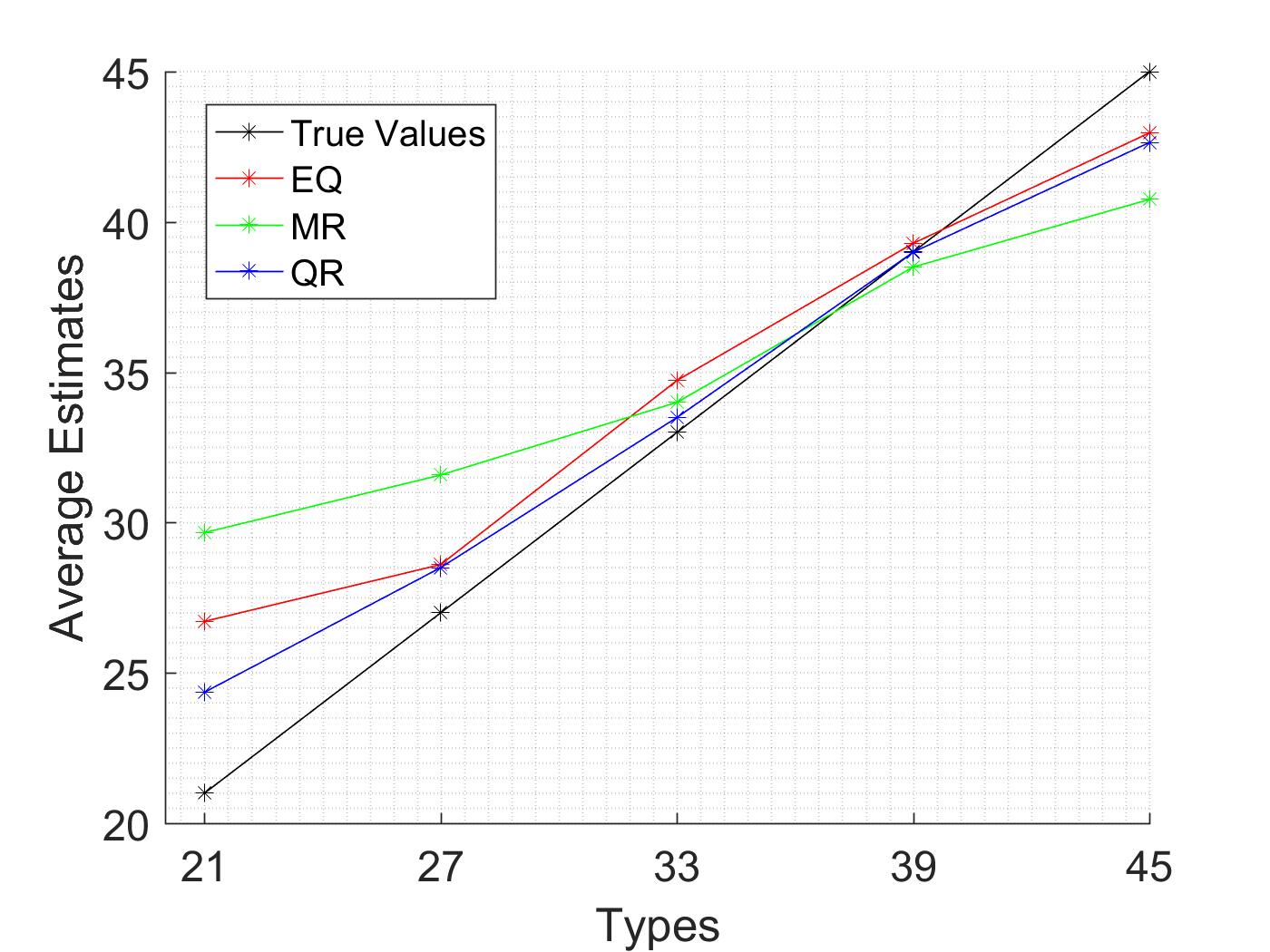}
  \caption{VCG estimates \label{fig:vcg-avg-ests}}
\end{subfigure}
\caption{Estimation results in VCG sessions: quantal-regret (QR) vs. min-regret (MR) vs. equilibrium-based (EQ) methods. 
(\ref{fig:vcg-bars-info}) and (\ref{fig:vcg-bars-type}) present the RMSE by value-information conditions and by types of players, respectively. 
(\ref{fig:vcg-avg-ests}) presents the average estimates of the three methods according to types of players alongside the true value for each type. 
 \label{fig:vcg-bars}}
\end{figure}

Quantal regret consistently outperforms the other two methods also when considering the estimation errors by information settings and by types of players. Figures \ref{fig:vcg-bars-info} and \ref{fig:vcg-bars-type} present the RMSE by the different experimental settings and Table \ref{tbl:ad-auction-full} (left column) provides the full estimation results by setting. 
First, Figure \ref{fig:vcg-bars-info} shows that both when players are given with their values (the GV setting) or when they have to deduce their own value (the DV setting), %in the deduced-value and in the given-value settings, 
the quantal regret method has lower RMSE than each of the other two methods, again by a large gap, and this was significant at the 5\% level for both settings (paired two-sided Wilcoxon signed rank tests, N=6 sessions).  
In addition, all three methods succeed somewhat better in estimating values in the given-value setting than in the deduced-value one, as could be expected, however these differences were not statistically significant for neither of the methods.

Second, also when considering each type separately, we found that the RMSE of the quantal regret method is consistently lower than the RMSE of the other methods, and it achieved much higher hit-rates than the others (see Figure \ref{fig:vcg-bars-type} and Table \ref{tbl:ad-auction-full}, 
except for type 33 for which the min-regret is better). 
In addition, as was found and discussed in depth in \cite{NN2017}, using the basic min-regret method there is a significant negative correlation between the estimation error and the player type ($\rho=-0.34$, $p<0.01$), indicating that it tends to perform better on the higher type players than on the lower types.
The quantal-regret method ``flattens'' the errors (due to a larger improvement on the two lowest types) and reduces this correlation to only $\rho=-0.12$ ($p=0.37$). 
Finally, Figure \ref{fig:vcg-avg-ests} plots the average estimates of the three methods alongside the true value for each player type. As can be seen, the three methods tend to err to the same direction, overestimating the values of the lower type players and underestimating the values of players of the highest type. %and the quanal regret estimates tend to be closer on average to the true values for each of the types. 

To conclude, in the VCG setting, the estimates obtained by the quantal regret method are consistently and significantly better than the estimates using the other two methods: it improves upon the min-regret method, as well as upon the standard method that relies on the specific and strong prediction of the truthful equilibrium for the VCG.

\subsubsection{Evaluating estimations in GSP auctions}

For the GSP auction the situation is much more complicated for equilibrium-based econometric methods, since there are no dominant strategies and there exist multiple equilibria.  There are two basic approaches in the literature for deducing bidders' valuations in GSP auctions, and we use them to compare the quantal regret performance. 
The first method was suggested in \cite{Varian2007} and is based on the assumption that the players reach the ``VCG-like'' equilibrium of the GSP, which is the equilibrium of the full-information one-shot game that gives the VCG-prices (we denote this method by ``EQ1''). 
The second method was suggested by \cite{Athey2010}, where bidders participate in a large number of auctions, and receive feedback that can vary from auction to auction, and the basic assumption is that each bidder is best-responding to the {\em distribution} that he faces (we denote this method by ``EQ2''). For more details regarding these two methods see Appendix \ref{app:gsp-methods}.

Table \ref{fig:bottom-auctions} presents the bottom line of the comparison results of the quantal regret method with the min-regret  method and the two ``classic'' equilibrium-based methods, over all 60 GSP bidders. As can be seen, %and similar to the case of VCG, 
the quantal regret method %performs much better than each of the other methods. 
outperforms the min-regret method, which in turn outperforms the two equilibrium-based methods. 
%Specifically, the RMSE of the QR method is 5.1, while the other methods have much higher RMSE, %of 8 using the MR method and of above 9.5 using the two classic methods, 
The RMSE of the QR method is far lower than the RMSE of the other methods, 
and the differences are statistically significant: the estimation errors obtained using the QR method are significantly lower than the errors obtained using each of the other three methods (paired two-sided Wilcoxon signed rank tests, N=12 sessions, $p<0.001$), with an average error of a less than half of that obtained using the two classic equilibrium-based methods. 
The advantage of the min-regret over EQ1 and EQ2 is also statistically significant over all GSP players ($p<0.01$).
\footnote{
This is different from the results in \cite{NN2017}, where MR was found to be significantly different only from the EQ1 method.  
%The results in the current paper are slightly different from those reported in \cite{NN2017} 
The difference is 
due to the use of the entire auction game (rather than using only the second half in \cite{NN2017}), and the use of the 
absolute estimation error (rather than using the relative error) which seems to help the MR method that tends to have higher error on the lower value bidders. 
In Section \ref{sec:robust-auctions} (Tables \ref{tbl:ad-auction-half} and \ref{tbl:ad-auction-rel-err}) we show that the new quantal regret method is completely robust to these variants. 
} 
Furthermore, quantal regret is the only method that achieves a reasonable hit-rate with 81.67\% of the estimates within a delta of 6 coins of the true value, while all other methods perform poorly and only manage to achieve hit-rates below 60\%.

\begin{figure} % slide 12
\begin{subfigure}{.32\textwidth} 
  \includegraphics[scale=0.13]{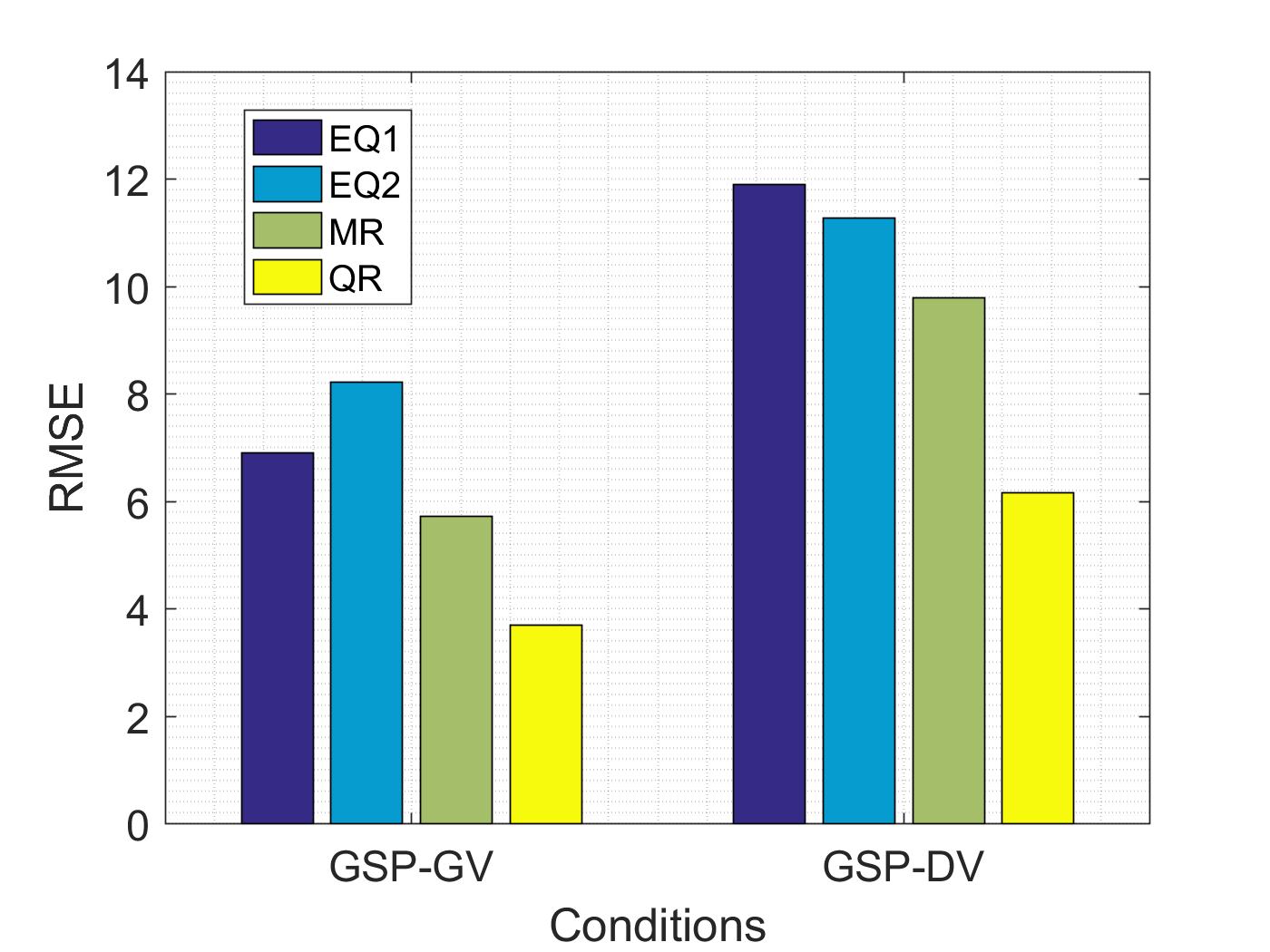}
  \caption{By information %conditions 
	\label{fig:gsp-bars-info}}
\end{subfigure}
\begin{subfigure}{.32\textwidth}
  \includegraphics[scale=0.13]{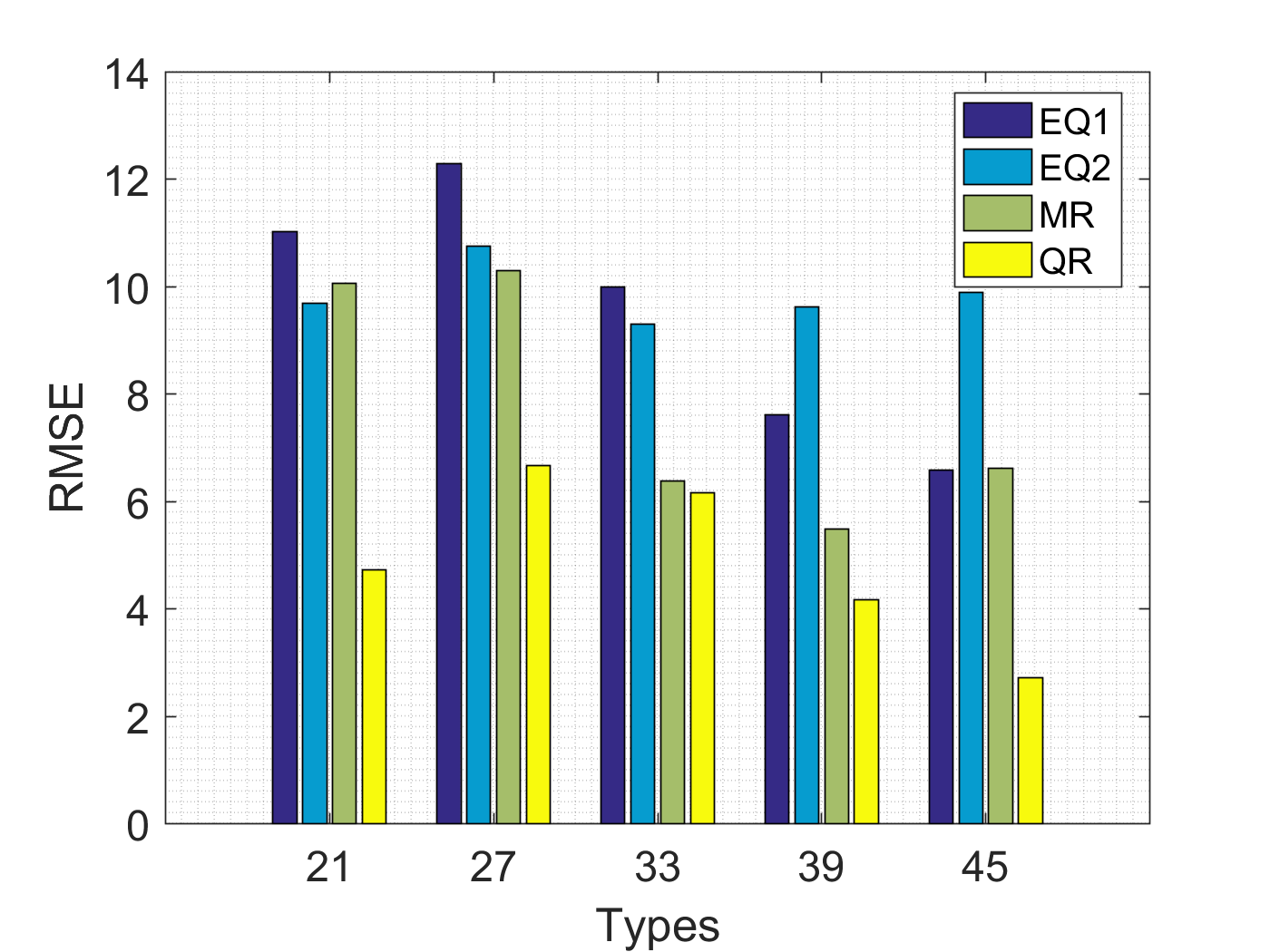}
  \caption{By player types \label{fig:gsp-bars-type}}
\end{subfigure}
\begin{subfigure}{.32\textwidth}
  \includegraphics[scale=0.13]{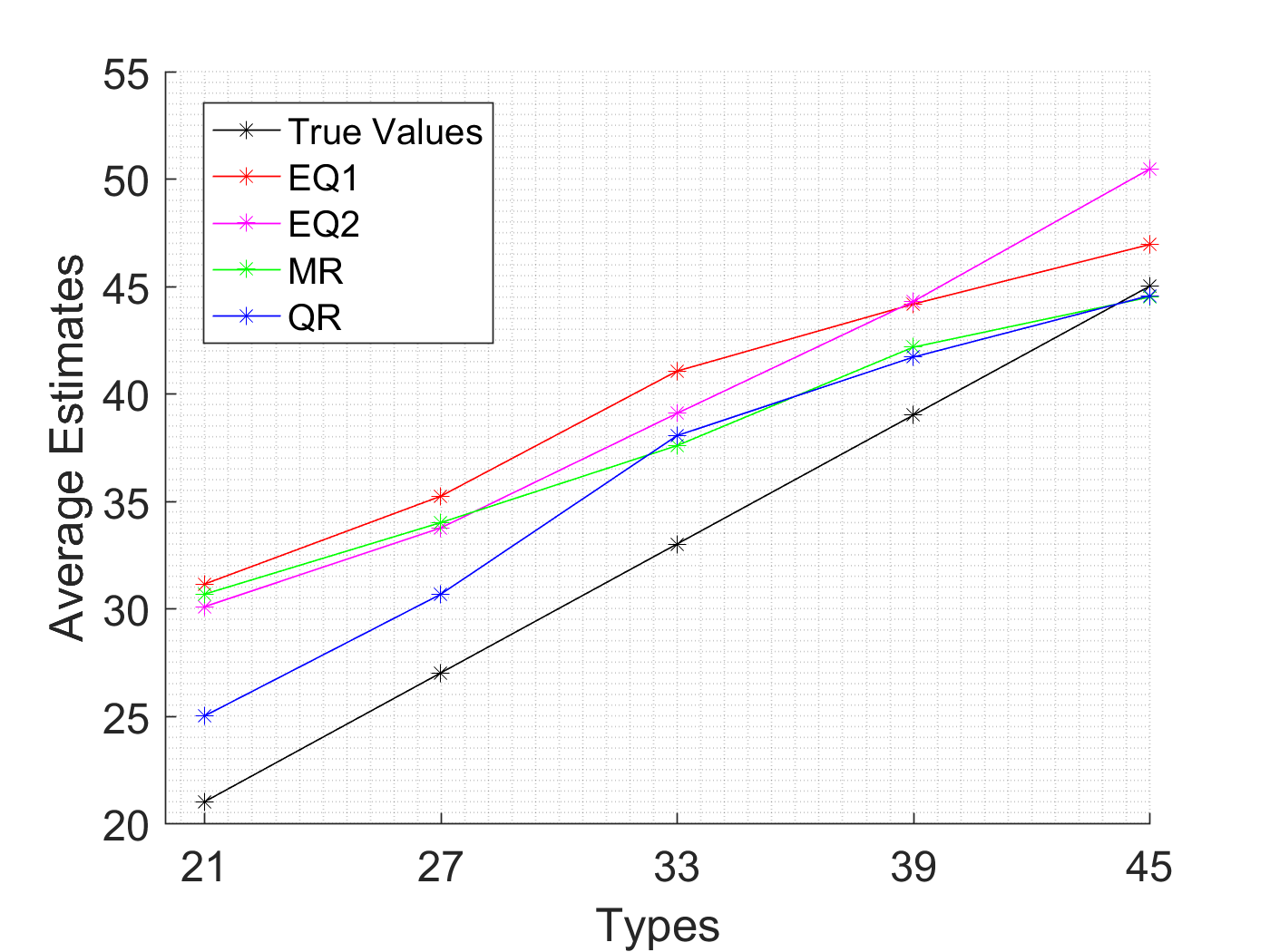}
  \caption{GSP estimates \label{fig:gsp-avg-ests}}
\end{subfigure}
\caption{Estimation results in GSP sessions: quantal-regret (QR) vs. min-regret (MR) vs. the two equilibrium-based methods (EQ1 of \cite{Varian2007} and EQ2 of \cite{Athey2010}). 
(\ref{fig:gsp-bars-info}) and (\ref{fig:gsp-bars-type}) present the RMSE by value-information conditions and by types of players, respectively. 
(\ref{fig:gsp-avg-ests}) presents the average estimates of the four methods according to types of players alongside the true value for each type. 
 \label{fig:gsp-bars}}
\end{figure}

As in the case of VCG, the advantage of the quantal regret method is robust to whether considering the estimation errors by information settings or by types of players, as can be seen in Figures \ref{fig:gsp-bars-info} and \ref{fig:gsp-bars-type} and in Table \ref{tbl:ad-auction-full} (right column). 
First, in both information settings GV and DV (Figure \ref{fig:gsp-bars-info}), the estimation errors using the QR method were significantly lower than the errors using each of the three other methods (paired two-sided Wilcoxon signed rank tests, N=6 sessions, $p<0.05$). 
In addition, the QR method as well as the other methods have higher error in the deduced-value setting than in the given-value setting, as could be expected, and these effects are statistically significant for the QR and the EQ1 methods (two-sample t-test, N=6 sessions, $p<0.05$ for QR and EQ1; testing for MR and EQ2  result in $p=0.06$ and $p=0.13$, respectively).

Second, the quantal regret method has much lower RMSE than the other methods also when looking at each type separately (see Figure \ref{fig:gsp-bars-type}), as well as much higher hit-rates (see Table \ref{tbl:ad-auction-full}). 
In addition, as was found and discussed in depth in \cite{NN2017}, we again find that the min-regret method succeeds better in estimating values of higher-type players than of the lower-types -- it has a significant negative correlation between the error and the player-type ($\rho=-0.31$, $p<0.02$),
and similar to the case of VCG above we find that the quantal regret method flattens the error and reduces this correlation to $\rho=-0.22$ ($p=0.09$). 
Finally, Figure \ref{fig:gsp-avg-ests} plots the average estimates of the four methods for each type, and shows that all methods err to the same direction relative to the true value. %The tendency of all methods to overestimate the values in GSP sessions indicate that the players bid higher than expected, and is consistent with the finding in \cite{NNY2014} that players in GSP auctions shaded their bids but not enough to reach the equilibrium prediction. 

To conclude, we find that also in GSP, the quantal regret method consistently and significantly improves upon the performance of the basic min-regret method, as well as upon the two ``classic'' equilibrium based methods that make stronger assumptions and are specific to the GSP auction rules.

\subsection{Further Robustness of the Results} \label{sec:robust-auctions}

\begin{table}
\centering
\resizebox{\textwidth}{!}{ %scale down table to the textwidth
\begin{tabular}{|l||l l l||l l l l|}
\hline
%\multicolumn{8}{|l|}{Over All Players} \\ \hline
&  \multicolumn{3}{|c||}{VCG Sessions} &  \multicolumn{4}{|c|}{GSP Sessions} \\ \hline
&  EQ & MR & QR & EQ1 & EQ2 & MR & QR \\ 
\hline
RMSE  & 6.51 & 6.01 & 3.45 & 8.00 & 7.96 & 6.60 & 4.46 \\ 
Avg Err  & 5.04 & 5.00 & 2.63 & 6.02 & 5.83 & 5.08 & 3.23\\
 $\pm 6$ Hit Rate & 61.67\% & 73.33\% & 88.33\% & 61.67\% & 56.67\% & 63.33\% & 91.67\% \\
\hline
\end{tabular}
}
\caption{Robustness to Auction Selections: Estimation results in the ad auction dataset, over all players for either GSP or VCG sessions, computed for the second half of the auction game (with 750 auctions) rather than for the entire auction game. \label{tbl:ad-auction-half}}
\end{table}

\begin{table}
\centering
\resizebox{\textwidth}{!}{ %scale down table to the textwidth
\begin{tabular}{|l||l l l||l l l l|}
\hline
%\multicolumn{8}{|l|}{Over All Players} \\ \hline
&  \multicolumn{3}{|c||}{VCG Sessions} &  \multicolumn{4}{|c|}{GSP Sessions} \\ \hline
&  EQ & MR & QR & EQ1 & EQ2 & MR & QR \\ 
\hline
RMSE & 22.4\% & 24.1\% & 14.3\%  & 35.6\% & 33.5\% & 30.1\% & 18.0\% \\ 
Avg Err  & 17.9\% & 18.1\% & 11.3\% & 26.8\% & 26.4\% & 22.1\% & 13.0\% \\
 $\pm 20\%$ Hit Rate & 63.33\% & 66.67\% & 85.00\%  & 48.33\% & 43.33\% & 56.67\% & 83.33\% \\
\hline
\end{tabular}
}
\caption{Robustness to Error Definition: Estimation results in the ad auction dataset, over all players for either GSP or VCG sessions, computed by  relative error (i.e., as percentages of the players' true value) rather than by absolute error.  \label{tbl:ad-auction-rel-err}}
\end{table}

In the previous section we showed how one implementation of the quantal regret method performs better than the other methods and that these results are robust to the different experimental settings. In this section we show the robustness of the results to other implementation variants. 

First, the results reported above were obtained using the regret aversion parameter $\lambda=1$, however in fact quantal regret outperforms the other methods for all reasonable values of $\lambda$. 
Figures \ref{fig:lambda-vcg} and \ref{fig:lambda-gsp} show the RMSE using the quantal regret method for players in the VCG and the GSP sessions, respectively, with different values of $\lambda$, compared with the min-regret method and the classic equilibrium-based methods. As can be seen, while the minimal RMSE of the quantal regret method is obtained around $\lambda=0.5$, it remains lower than the other methods and (as expected) approaches the RMSE of the MR method as $\lambda$ grows large. The QR error is large for $\lambda \approx 0$, where it effectively ignores the regrets and simply averages all values according to the prior distribution (which is the uniform distribution in our case).

Second, since the players in the ad auction experiment had only partial information about the game, it might be that 
the equilibrium-based methods which assume stability could have won had we performed the evaluation after the game has stabilized (rather than based on the entire auction game).  
However, Table \ref{tbl:ad-auction-half} shows that the quantal regret method outperforms the other methods, according to all three criteria, also when using only the second half of the auction game (750 auctions), excluding the first half as an initial learning phase (as was done in \cite{NN2017}). 
Finally, for consistency with \cite{NN2017}, 
we report that comparing the quantal regret with the other methods based on the relative error (i.e., by considering the error relative to the true value) had lead to similar results and even slightly increased the gap between the quantal regret and the other methods, as can be seen in Table \ref{tbl:ad-auction-rel-err}.

\section*{ACKNOWLEDGMENTS}
We would like to thank Thorsten Chmura for generously sharing with us his 2x2 game dataset.

This research is supported by ISF grant 1435/14 administered by the Israeli Academy of Sciences, by
Israel-USA Bi-national Science Foundation (BSF) grant number 2014389, and 
by the Adams Fellowship Program of the Israel Academy of Sciences and Humanities. 
%end ACKNOWLEDGMENTS

\bibliographystyle{apalike} % abbrv}
\bibliography{regret-bib}

\begin{thebibliography}{}

\bibitem[Arora et~al., 2012]{Arora2012}
Arora, S., Hazan, E., and Kale, S. (2012).
\newblock The multiplicative weights update method: a meta-algorithm and
  applications.
\newblock {\em Theory of Computing}, 8(1):121--164.

\bibitem[Athey and Nekipelov, 2010]{Athey2010}
Athey, S. and Nekipelov, D. (2010).
\newblock A structural model of sponsored search advertising auctions.

\bibitem[Aumann, 1974]{Corr1974}
Aumann, R.~J. (1974).
\newblock Subjectivity and correlation in randomized strategies.
\newblock {\em Journal of Mathematical Economics}, 1(1):67 -- 96.

\bibitem[Blum and Mansour, 2007]{BM2007}
Blum, A. and Mansour, Y. (2007).
\newblock Learning, regret minimization, and equilibria.
\newblock In {\em Algorithmic Game Theory}. Cambridge: Cambridge University
  Press.

\bibitem[Edelman et~al., 2007]{EOS2007}
Edelman, B., Ostrovsky, M., and Schwarz, M. (2007).
\newblock Internet advertising and the generalized second-price auction:
  Selling billions of dollars worth of keywords.
\newblock {\em American Economic Review}, 97(1):242--259.

\bibitem[Hart and Mas-Colell, 2000]{Hart2000}
Hart, S. and Mas-Colell, A. (2000).
\newblock A simple adaptive procedure leading to correlated equilibrium.
\newblock {\em Econometrica}, 68(5):1127--1150.

\bibitem[McKelvey and Palfrey, 1995]{QRE1995}
McKelvey, R.~D. and Palfrey, T.~R. (1995).
\newblock Quantal response equilibria for normal form games.
\newblock {\em Games and Economic Behavior}, 10(1):6 -- 38.

\bibitem[Nekipelov et~al., 2015]{Nekipelov2015}
Nekipelov, D., Syrgkanis, V., and Tardos, E. (2015).
\newblock Econometrics for learning agents.
\newblock In {\em Proceedings of the Sixteenth ACM Conference on Economics and
  Computation}, EC '15, pages 1--18, New York, NY, USA. ACM.

\bibitem[Nisan and Noti, 2017]{NN2017}
Nisan, N. and Noti, G. (2017).
\newblock An experimental evaluation of regret-based econometrics.
\newblock {\em Proceedings of the 26th International Conference on World Wide
  Web}.

\bibitem[Noti et~al., 2014]{NNY2014}
Noti, G., Nisan, N., and Yaniv, I. (2014).
\newblock An experimental evaluation of bidders' behavior in ad auctions.
\newblock In {\em Proceedings of the 23rd International Conference on World
  Wide Web}, WWW '14, pages 619--630, New York, NY, USA. ACM.

\bibitem[Selten and Chmura, 2008]{Selten2008}
Selten, R. and Chmura, T. (2008).
\newblock Stationary concepts for experimental 2x2-games.
\newblock {\em American Economic Review}, 98(3):938--66.

\bibitem[Varian, 2007]{Varian2007}
Varian, H.~R. (2007).
\newblock Position auctions.
\newblock {\em International Journal of Industrial Organization},
  25(6):1163--1178.

\bibitem[Young, 2004]{Coarse}
Young, H.~P. (2004).
\newblock {\em Strategic learning and its limits}.
\newblock OUP Oxford.

\end{thebibliography}

\newpage

\appendix

\section*{APPENDICES}

\FloatBarrier

\section{The Min-Regret Method Using the Relative Regret} \label{app:mr-rel-regret}
%rel-regret.tex

\begin{table}[h]
\centering
\begin{subtable}{.5\textwidth}
\centering

\resizebox{\textwidth}{!}{ %scale down table to the textwidth
\begin{tabular}{|l|l|l|l|}
\hline
\multicolumn{4}{|c|}{2x2 Games -- Over All Sessions} \\ \hline
	& EQ & MR & MR-REL \\ \hline
RMSE  & 3.41 & 3.25 & 3.25 \\
Average Error  & 2.99 & 2.84 & 2.85 \\
$\pm 3$ Hit Rate  & 68.87\% & 75.00\% & 75.00\% \\
\hline
\end{tabular}
}
\caption{\label{fig:bottom-games-rel}}
\end{subtable} 

\begin{subtable}{1.0\textwidth}
\centering

\resizebox{\textwidth}{!}{ %scale down table to the textwidth
\begin{tabular}{|l|l|l|l|l|l|l|l|}
\hline
& \multicolumn{3}{|c|}{Ad auctions -- VCG Sessions}  & \multicolumn{4 }{|c|}{Ad auctions -- GSP Sessions} \\ \hline
	& EQ & MR & MR-REL & EQ1 & EQ2 & MR & MR-REL \\ \hline
RMSE  & 6.46 & 6.26 & 7.15 						& 9.73 & 9.87 & 8.02 & 11.06 \\
Average Error  & 5.38 & 5.13 &  5.95 		 	& 7.74 & 8.02 & 6.32 & 9.37  \\
$\pm 6$ Hit Rate  & 61.67\% & 63.33\% & 56.67\%			& 48.33\% & 41.67\% & 56.67\% &  43.33\% \\ 
\hline
\end{tabular}
}

\caption{\label{fig:bottom-auctions-rel}}
\end{subtable}

\caption{Estimation results of the min-regret method compared with the equilibrium-based methods, in the 2x2 game dataset (\ref{fig:bottom-games-rel}) and in the ad auction dataset (\ref{fig:bottom-auctions-rel}). 
The min-regret method is based on either the relative-regret (MR-REL) as was originally suggested by \cite{Nekipelov2015} (i.e., the regret is relative to the optimal outcome on fixed action), or on the absolute-regret (MR) as is evaluated in the current paper. As can be seen, in the ad auction setting the min-regret performs much better using absolute-regret than using relative-regret, and in this paper we use the more precise method as a tougher benchmark for comparison. }
\end{table}

\newpage

\FloatBarrier

\section{Game Utilities in the 2x2 Game Dataset} \label{app:games}

\begin{figure}[h]
\begin{center}
\includegraphics[scale=1.41]{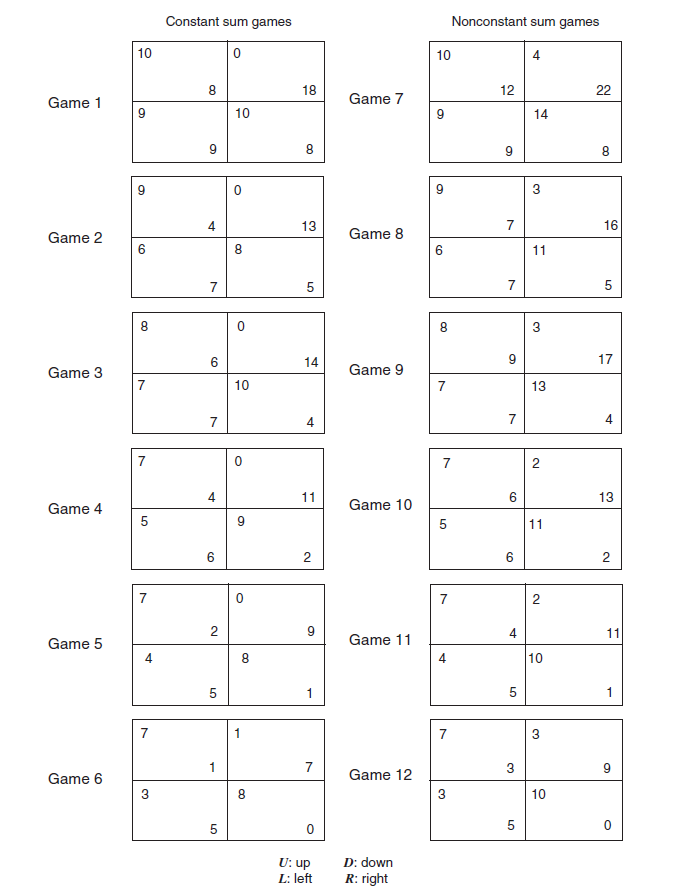}
\end{center}
\caption{Experimentally investigated games from \cite{Selten2008}. Games 1-6 are constant sum games and games 7-12 are non-constant sum games. The upper-left and the lower-right corners in each cell are the payoffs of the row and the column players, respectively.
 \label{fig:games-util}}
\end{figure}

\FloatBarrier

\section{Robustness of the results in the 2x2 Game Dataset -- Additional Tables} \label{app:games-robust}
%app-games-robust.tex

\subsection{Robustness to the Grid Selection}
\FloatBarrier

Table \ref{tbl:grid} shows the estimation results for the 2x2 game dataset, when varying the resolution of the grid being used for the regret-based methods in section \ref{sec:games-results}. As can be seen, increasing the grid resolution rarely affects the results of the quantal regret method, and it outperforms the two other methods in all cases. In contrast, the min-regret results become closer and closer to the EQ results, as expected, since the estimates of the MR estimator that allows real values would be identical to those of the EQ method (as explained in Section \ref{sec:intro}).

\begin{table}[h]
\centering
\begin{tabular}{|l|l|l l l|}
\hline
%& \multicolumn{3}{|l|}{72 Constant Sum Sessions} & \multicolumn{3}{|l|}{36 Non-Constant Sum Sessions} \\ \hline
Grid & & EQ & MR & QR  \\ \hline
1  & RMSE  & 3.413 & 3.254 & 2.290 \\ 
& Average Error  & 2.986 & 2.844 & 2.041 \\
& $\pm 3$ Hit Rate & 68.87\% & 75.00\% & 81.60\% \\
\hline
0.1  & RMSE  & 3.413 & 3.400 & 2.316 \\ 
& Average Error & 2.986 & 2.974 & 2.081  \\
& $\pm 3$ Hit Rate & 68.87\% & 69.10\% & 81.13\% \\
\hline
0.01  & RMSE & 3.413 & 3.411 & 2.319 \\ 
& Average Error & 2.986 & 2.984 & 2.086 \\
& $\pm 3$ Hit Rate & 68.87\% & 69.10\% & 81.13\% \\
\hline
0.001  & RMSE & 3.413 & 3.413 & 2.319 \\ 
& Average Error & 2.986 & 2.986 & 2.086 \\
& $\pm 3$ Hit Rate & 68.87\% & 68.87\% & 81.13\% \\
\hline
\end{tabular}
\caption{Estimation results over all 108 sessions of the 2x2 game dataset, for different resolution of the grid being used for the regret calculation. 
\label{tbl:grid}}
\end{table}

\FloatBarrier

\subsection{Special Treatment for the Constant Sum Games}

Table \ref{tbl:levels-cs} shows the estimation results for the three estimation methods for the constant sum games, when the estimation takes into account the constant sum property of the game, i.e., that for each of the four game outcomes the payoffs of the two players sum up to some constant $C$. % (as explained in Section \ref{sec:games-robust}). 
In this case, the estimation is of the 4 parameters defining the constant sum game, by aggregating the results of the two players in the session level, using the parameter range $[0,C]$ (where $C$ is the game constant). 
As can be seen, the aggregation improves the estimation results for all three methods (QR, EQ, and MR), as could be expected. The improvement for the MR method is sharper and in the aggregated case the MR outperforms the QR method.

\begin{table}[h]
\centering
\resizebox{\textwidth}{!}{ %scale down table to the textwidth
\begin{tabular}{|l|l|l l l|}
\hline
\multicolumn{2}{|l|}{Estimation Procedure For Constant Sum Games} & EQ & MR & QR  \\ \hline
Session-level estimate & RMSE & 3.06 & 2.90 & 2.33 \\ 
(8 parameters) &  Average Error  & 2.70 & 2.55 & 2.17 \\
&  $\pm 3$ Hit Rate & 73.78\% & 78.65\% & 80.56\%  \\
\hline
Session-level estimates aggregated & RMSE & 2.35 & 1.66 & 1.94 \\ 
(4 parameters) &  Average Error & 2.06 & 1.45 & 1.82  \\
&  $\pm 3$ Hit Rate & 75.69\% & 93.40\% & 85.07\%  \\
\hline
\end{tabular}
}
\caption{Estimation results for the constant sum games of the 2x2 game dataset, 
by either estimating only 4 parameters or estimating the 8 parameters separately. 
In both cases, the estimation is in the parameter range $[0,C]$, where $C$ is the game constant.  
\label{tbl:levels-cs}}
\end{table}

\FloatBarrier

\newpage

\subsection{Estimation Results by Game}

\begin{table}[h]
\centering
\resizebox{\textwidth}{!}{ %scale down table to the textwidth
\begin{tabular}{|l||l|l l l||l|l l l|}
\hline
& & \multicolumn{3}{|l||}{Constant Sum Games} &  & \multicolumn{3}{|l|}{Non-Constant Sum Games} \\ \hline
&  Game & EQ & MR & QR & Game & EQ & MR & QR \\ \hline
RMSE & 1  & 4.92 & 4.65 & 3.01 &  7   & 5.90 & 5.47 & 4.02 \\ 
Avg Err & & 4.84 & 4.57 & 2.95 &     & 5.89 & 5.46 & 3.98 \\
 $\pm 3$ Hit Rate & & 50.00\% & 52.08\% & 65.63\% &   & 33.33\% & 33.33\% & 41.67\% \\
\hline
RMSE & 2  & 3.52 & 3.36 & 2.49 &  8   & 2.94 & 2.92 & 2.49  \\ 
Avg Err  & & 3.38 & 3.21 & 2.36 &   & 2.82 & 2.79 & 2.32 \\
$\pm 3$ Hit Rate  & & 54.17\% & 66.67\% & 71.88\% &   & 58.33\% & 70.83\% & 83.33\% \\
\hline
RMSE & 3  & 4.36 & 4.15 & 2.08 &  9   & 3.65 & 3.55 & 3.04  \\ 
Avg Err  & & 4.25 & 4.04 & 2.06 &   & 3.46 & 3.37 & 2.84 \\
$\pm 3$ Hit Rate  & & 54.17\% & 60.42\% & 83.33\% &   & 58.33\% & 66.67\% & 66.67\% \\
\hline
RMSE & 4  & 2.71 & 2.59 & 1.66 &  10   & 2.80 & 2.70 & 2.35  \\ 
Avg Err  & & 2.65 & 2.52 & 1.62 &   & 2.70 & 2.61 & 2.32 \\
$\pm 3$ Hit Rate  & & 66.67\% & 77.08\% & 88.54\% &    & 66.67\% & 79.17\% & 72.92\% \\
\hline
RMSE & 5  & 2.54 & 2.50 & 2.01 &  11   & 1.83 & 1.72 & 1.20  \\ 
Avg Err  & & 2.07 & 1.99 & 1.66 &  & 1.74 & 1.64 & 1.13 \\
$\pm 3$ Hit Rate  & & 90.63\% & 93.75\% & 92.71\% &   & 91.67\% & 100.00\% & 100.00\% \\
\hline
RMSE & 6  & 0.99 & 1.00 & 1.02 &  12   & 1.07 & 1.04 & 0.97  \\ 
Avg Err  & & 0.90 & 0.84 & 0.95 &   & 0.95 & 0.96 & 0.93 \\
$\pm 3$ Hit Rate  & & 100.00\% & 100.00\% & 100.00\% &   & 100.00\% & 100.00\% & 100.00\% \\
\hline
RMSE & All & 3.42 & 3.27 & 2.14 &  All   & 3.39 & 3.23 & 2.57   \\ 
Avg Err & 1-6 & 3.02 & 2.86 & 1.93  &  7-12  & 2.93 & 2.81 & 2.25  \\
$\pm 3$ Hit Rate & & 69.27\% & 75.00\% & 83.68\% &   & 68.06\% & 75.00\% & 77.43\% \\
\hline 
\end{tabular}
}
\caption{Estimation results in the 2x2 game dataset, for each of the 12 games separately, and for all constant sum games (games 1-6 with 12 sessions each) and all non-constant sum games (games 7-12 with 6 sessions each). \label{tbl:rmse-12-games}}
\end{table}

\FloatBarrier

\newpage

\section{Game Interface in the Ad Auction Experiment} \label{app:screenshot}

%Figure \ref{fig:screenshot} presents a screen shot of the user interface that was used in the ad auction experiment of \cite{NNY2014}. 

\begin{figure}[h]
\centerline{\includegraphics[scale=0.38]{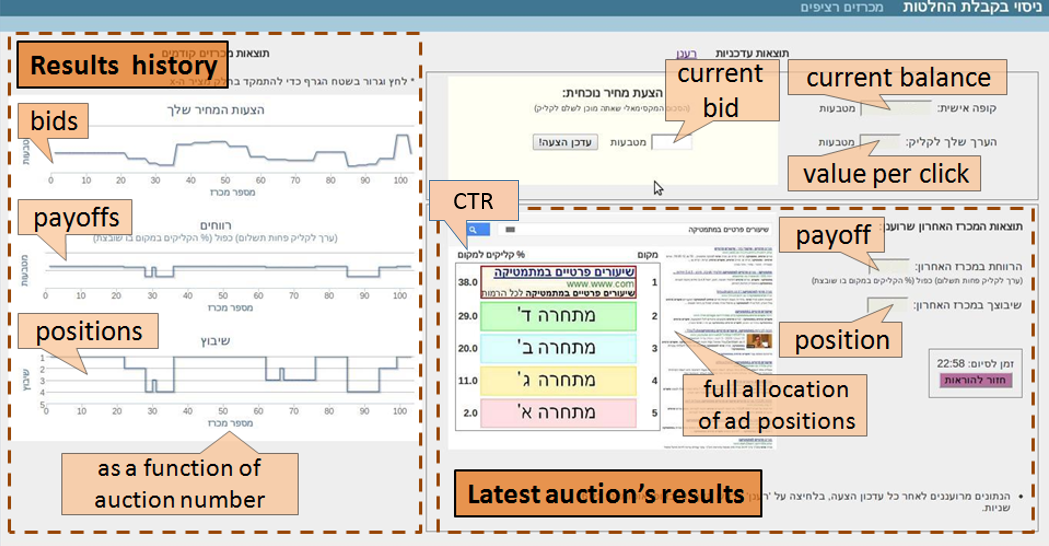}}
\caption{Screen shot of the user interface used in the ad auction experiment of \cite{NNY2014}. \label{fig:screenshot} }
\end{figure}

\FloatBarrier

\section{Summary of Existing Estimation Methods for GSP} \label{app:gsp-methods}
%gsp-methods.tex

\subsection{The VCG-like-NE Method (Denoted by ``EQ1'')}

In \cite{Varian2007} it is suggested that the players should reach the equilibrium of the full-information one-shot GSP game that gives the VCG-prices (hence the ``VCG-like'' equilibrium).  Assuming that this is indeed the case, then at each time step $t$ in a sequence of auctions, one may deduce values $\hat{v}_i^t$ for all players $i$ from the actual bids $b_i^t$, such that the bids are this VCG-like equilibrium of these deduced values. The final estimate is then the average of these $\hat{v}_i^t$.   Some complications arise when this is attempted on real data since it is often the case that the bids do not correspond to an equilibrium of any tuple of values. In these cases we follow \cite{Varian2007} and perturb the bid observations in the minimal possible way so as to satisfy the equilibrium constraints, and set the final estimates to the perturbed values.\footnote{In fact, only 13.3\% of the auctions were consistent with the equilibrium inequalities without perturbing their data. However, similar to \cite{Varian2007}, we observed that the required perturbations were relatively small.} These and other complications of this method are discussed in \cite{Varian2007} and in the full version of \cite{NN2017}.

\subsection{The Best-Response Method (Denoted by ``EQ2'')}

A %more complex 
second method was suggested by \cite{Athey2010} where bidders participate in a large number of auctions, and receive feedback that can vary from auction to auction. The basic assumption is that each bidder is best-responding to the {\em distribution} that he faces (by placing a single bid).
%Specifically, this ``Best-Response'' method assumes that players best-respond by taking into account the following two functions of their bid: the expected CTR of the position they win and their expected payment.  
Specifically, given a sequence of auctions, define functions $Q_i(b_i)$ and $TE_i(b_i)$ as the expected CTR and the expected total expenditure, respectively, of bidder $i$ by bidding $b_i$. Thus, his expected utility with valuation $v$ is $Q_i(b_i) \cdot v - TE_i(b_i)$. Against {\em smooth} distributions the best bid would be a strictly increasing function of the value. %so the single value that corresponds to the actual bid may be served as the value estimate.
In these cases, the valuation of bidder $i$ who maximizes his expected utility by bidding $b_i$ can be recovered using the first-order condition by $\hat{v}_i = \frac{\partial TE_i(b_i)/\partial b_i}{\partial Q_i(b_i)/\partial b_i}$. When applying this method to actual data complications arise, and there are many possible implementations. We have tested several implementation variants,  and in the implementation we chose %(referred to as the ``Best-Response'' method), 
(denoted by ``EQ2'' in this paper), we used the average bid that a bidder played as his best-response to the distribution of the bids of the others (since bids were not constant), and found the value by optimizing directly using grid search (since the empirical derivatives had their own complications). Details of implementation and complications of this method are discussed in \cite{Athey2010} and in the full version of \cite{NN2017}.

\FloatBarrier

\newpage

\section{Robustness of the Results in the Ad Auction Dataset -- Additional Tables} \label{app:games-robust}
%app-ads-robust.tex

\begin{table}[h]
\centering
\resizebox{\textwidth}{!}{ %scale down table to the textwidth
\begin{tabular}{|l|l||l l l||l l l l|}
\hline
& & \multicolumn{3}{|c||}{VCG Sessions} &  \multicolumn{4}{|c|}{GSP Sessions} \\ \hline
&  Setting & EQ & MR & QR & EQ1 & EQ2 & MR & QR \\ 
\hline
\hline
RMSE & All Players & 6.46 & 6.26 & 4.22 & 9.73 & 9.87 & 8.02 & 5.09  \\ 
Avg Err & & 5.38 & 5.13 & 3.42 & 7.74 & 8.02 & 6.32 & 3.85 \\
 $\pm 6$ Hit Rate & & 61.67\% & 63.33\% & 81.67\% & 48.33\% & 41.67\% & 56.67\% & 81.67\% \\
\hline
\hline
RMSE & GV & 5.95 & 6.05 & 3.98 			& 6.91 & 8.22 & 5.72 & 3.71 \\ 
Avg Err & & 5.04 & 5.03 & 3.07 			& 5.46 & 6.92 & 4.90 & 2.81 \\
$\pm 6$ Hit Rate & & 63.33\% & 63.33\% & 76.67\% 		& 70.00\% & 50.00\% & 70.00\% & 90.00\% \\
\hline
RMSE & DV & 6.93 & 6.47 & 4.44 			& 11.91 & 11.28 & 9.80 & 6.17 \\ 
Avg Err & & 5.72 & 5.23 & 3.77 			& 10.01 & 9.12 & 7.73 & 4.88 \\
 $\pm 6$ Hit Rate & & 60.00\% & 63.33\% & 86.67\% 		& 26.67\% & 33.33\% & 43.33\% & 73.33\% \\
\hline
\hline
RMSE & Type 21 & 7.64 & 9.31 & 4.16 		& 11.03 & 9.70 & 10.06 & 4.72 \\ 
Avg Err & & 7.15 & 8.67 & 3.77 					& 10.14 & 9.08 & 9.67 & 4.14 \\
$\pm 6$ Hit Rate & & 41.67\% & 25.00\% & 91.67\%  & 16.67\% & 16.67\% & 16.67\% & 75.00\% \\
\hline
RMSE & Type 27 & 4.32 & 5.61 & 3.84			 & 12.29 & 10.76 & 10.30 & 6.66 \\ 
Avg Err & & 3.50 & 4.92 & 2.87 						& 9.20 & 7.83 & 7.00 & 4.51	 \\
 $\pm 6$ Hit Rate & & 75.00\% & 66.67\% & 91.67\% 		& 50.00\% & 50.00\% & 58.33\% & 75.00\% \\
\hline
RMSE & Type 33 & 7.30 & 4.90 & 5.72 		& 10.00 & 9.31 & 6.38 & 6.16 \\ 
Avg Err &  & 6.20 & 4.17 & 5.04 				& 8.19 & 6.92 & 4.92 & 5.05 \\
 $\pm 6$ Hit Rate & & 58.33\% & 75.00\% & 58.33\% 		& 41.67\% & 66.67\% & 75.00\% & 75.00\% \\
\hline
RMSE & Type 39 & 5.60 & 3.19 & 3.07 		& 7.61 & 9.63 & 5.49 & 4.16 \\ 
Avg Err & & 4.12 & 2.17 & 2.45 					& 6.05 & 7.46 & 4.67 & 3.43 \\
 $\pm 6$ Hit Rate & & 75.00\% & 91.67\% & 91.67\% 			& 58.33\% & 41.67\% & 66.67\% & 83.33\% \\
\hline
RMSE & Type 45 & 6.84 & 6.61 & 3.84 		& 6.58 & 9.90 & 6.62 & 2.72 \\ 
Avg Err &  & 5.94 & 5.75 & 2.96 				& 5.11 & 8.79 & 5.33 & 2.10 \\
 $\pm 6$ Hit Rate &  & 58.33\% & 58.33\% & 75.00\% 		& 75.00\% & 33.33\% & 66.67\% & 100.00\% \\
\hline
\end{tabular}
}
\caption{Estimation results in the ad auction dataset, for either GSP or VCG sessions, and either over all players or according to player types or according to value-information conditions (Given-Value and Deduced-Value). \label{tbl:ad-auction-full}}
\end{table}

%\listoftodos

\end{document}